\documentclass[reqno,onefignum,onetabnum]{siamart190516}

\usepackage{graphicx}
\usepackage{subcaption}
\usepackage{float}
\usepackage{epstopdf}

\usepackage{xcolor}

\usepackage{chngcntr}
\counterwithout{figure}{section}
\counterwithout{equation}{section}

\newcommand{\pd}{\partial}

\DeclareMathOperator{\sign}{sign}

\makeatletter
\renewcommand*\env@matrix[1][\arraystretch]{%
  \edef\arraystretch{#1}%
  \hskip -\arraycolsep
  \let\@ifnextchar\new@ifnextchar
  \array{*\c@MaxMatrixCols c}}
\makeatother

\usepackage{lipsum}
\usepackage{amsfonts}
\usepackage{graphicx}
\usepackage{epstopdf}
\usepackage{algorithmic}
\ifpdf
  \DeclareGraphicsExtensions{.eps,.pdf,.png,.jpg}
\else
  \DeclareGraphicsExtensions{.eps}
\fi

\newcommand{\mc}{\mathcal }

\newsiamremark{remark}{Remark}
\newsiamremark{hypothesis}{Hypothesis}
\crefname{hypothesis}{Hypothesis}{Hypotheses}
\newsiamthm{claim}{Claim}

\headers{Stroboscopic motion reversals in delay-coupled neural fields}{N. Parks and Z.P. Kilpatrick}

\title{Stroboscopic motion reversals in delay-coupled neural fields \thanks{Submitted to the editors DATE.
\funding{This work was funded by NSF DMS-2207700 and NIH BRAIN 1R01EB029847-01.}}}

\author{Noah Parks\thanks{Department of Applied Mathematics, University of Colorado Boulder, Boulder, CO 
  (\email{noah.parks@colorado.edu}, \email{zpkilpat@colorado.edu})}
\and Zachary P Kilpatrick\footnotemark[2]
}

\usepackage{amsopn}

\usepackage{xr}
\makeatletter
\newcommand*{\addFileDependency}[1]{
  \typeout{(#1)}
  \@addtofilelist{#1}
  \IfFileExists{#1}{}{\typeout{No file #1.}}
}
\makeatother

\begin{document}

\maketitle
\begin{abstract}
Visual illusions provide a window into the mechanisms underlying visual processing, and dynamical neural circuit models offer a natural framework for proposing and testing theories of their emergence. We propose and analyze a delay-coupled neural field model that explains stroboscopic percepts arising from the subsampling of a moving, often rotating, stimulus, such as the wagon-wheel illusion. Motivated by the role of activity propagation delays in shaping visual percepts, we study neural fields with both uniform and spatially dependent delays, representing the finite time required for signals to travel along axonal projections. Each module is organized as a ring of neurons encoding angular preference, with instantaneous local coupling and delayed long-range coupling strongest between neurons with similar preference. We show that delays generate a family of coexisting traveling bump solutions with distinct, quantized propagation speeds. Using interface-based asymptotic methods, we reduce the neural field dynamics to a low-dimensional system of coupled delay differential equations, enabling a detailed analysis of speed selection, stability, entrainment, and state transitions. Regularly pulsed inputs induce transitions between distinct speed states, including motion opposite to the forcing direction, capturing key features of visual aliasing and stroboscopic motion reversal. These results demonstrate how delayed neural interactions organize perception into discrete dynamical states and provide a mechanistic explanation for stroboscopic visual illusions.
\end{abstract}

\begin{keywords}
  neural fields, delay differential equations, bump attractors, interface methods
\end{keywords}

\begin{AMS}
92B20, 45G10, 34K13, 37N25
\end{AMS}


\section{Introduction}
\label{sec:Introduction}

In old films, rapidly spinning spoked wheels can appear to rotate differently from their true rotation—a striking phenomenon known as the wagon wheel illusion. This effect arises from a mismatch between the wheel’s rotation rate and the camera’s frame rate, leading to a misperception of motion. The illusion was first documented in the late 19th century by Étienne-Jules Marey, whose chronophotographic studies revealed how sampling continuous motion in discrete frames can produce unexpected visual artifacts~\cite{marey1894mouvement}. Decades later, Adelson and Bergen formalized this observation, showing how such illusions emerge from the visual system’s interpretation of temporally sampled motion~\cite{adelson1985spatiotemporal}.

Surprisingly, a similar effect can occur even under continuous illumination~\cite{schouten1967,purves1996} as when observing the hub spokes of a car wheel or the blades of an airplane propeller. While the mechanism is less well understood, key differences include a delayed onset and the absence of any stationary percept~\cite{andrews2005}. One influential account interprets this as temporal aliasing caused not by an external shutter but by the brain’s own rhythmic sampling of visual input. In this view, perception unfolds in discrete cycles—an intrinsic “shutter speed” near the alpha band (10–13 Hz)—so that continuous motion is parsed into snapshots that can mimic the stroboscopic effect~\cite{vanrullen2005attention,vanrullen2016perceptual}.

Several theoretical frameworks have been proposed to account for wagon wheel phenomena, including correlation-based analyses of successive frames~\cite{adelson1985spatiotemporal,weiss2002}, recurrent cortical network models with excitatory--inhibitory feedback~\cite{glass2009}, and oscillator-based accounts emphasizing phase locking~\cite{martineau2009}. While these approaches capture important aspects of the stroboscopic illusion, they either predict stationary percepts that are not observed under continuous illumination or treat the effect as evidence against discrete temporal sampling, rather than providing a mechanistic explanation for its emergence. As a result, the continuous Wagon Wheel Illusion remains theoretically unresolved, motivating the present study.

This illusion likely reflects neural dynamics in early visual cortex, where motion signals are integrated and interpreted over time~\cite{baker1985,grzywacz1990}. The brain does not passively register motion; rather, perception emerges from structured activity in feature-selective neural populations, including those tuned to orientation and direction. Such dynamics provide a natural entry point for modeling the mechanisms underlying the continuous wagon wheel illusion.

Neurons in primary visual cortex (V1) exhibit orientation selectivity, responding most strongly when a stimulus aligns with a preferred angle~\cite{hubel1959}. These cells are spatially organized into orientation columns~\cite{blasdel1986}, creating a continuous map of orientation preferences across cortex. This functional architecture, together with the known interactions among orientation-selective populations, motivates the use of neural field models in which orientation is represented on a continuous ring.

In line with previous work, we adopt a neural field model to investigate the neural response to visual stimuli. Neural fields consist of integrodifferential equations to model the evolution of a continuous “activity density” of cortex tissue, rather than considering individual neurons as discrete points~\cite{wilson1973mathematical}. Such models typically describe bulk neural activity as decaying to rest unless provided with input or recurrent drive from an integral coupling term that encodes the strength of connections between spatially distinct neural populations. Following previous work on early visual processing, we assume that neurons are strongly connected to others with similar feature (orientation) preferences~\cite{benyishai1997}. Such network organization for orientation columns generates spatially coherent neural activity patterns~\cite{ermentrout1998neural}.

A central feature of these models and our work is the existence of ``bump” solutions, representing regions of localized neural activity that can be stabilized by a combination of short-range excitation and long-range inhibition~\cite{amari1977}. The centroid of the bump, which coincides with the location of its peak, represents the orientation preference of the neurons with the highest activity, while neurons with similar preferences experience similar activity. Hence in the context of a neural population receptive to rotational motion, the centroid of the bump may be interpreted as the perceived angular position of the stimulus. A traveling bump is then indicative of a rotating stimulus, with the direction of motion of the bump matching the direction of rotation. Such models have been used to describe representations of internal variables like head direction~\cite{zhang1996,kutschireiter2023bayesian} and external variables like visual stimulus location or orientation~\cite{camperi1998model,kilpatrick2013wandering,wimmer2014bump}.

Most bump attractor models focus on a single orientation ring, or on asymmetrically coupled rings acting as one unit~\cite{amari1977,zhang1996,benyishai1997}, yet motion representations in cortex are distributed and often redundant across areas~\cite{huk2001,cohen2010}. For instance, direction selectivity has long been linked to delay lines or temporally staggered inputs~\cite{reichardt1961,adelson1985spatiotemporal}, consistent with measured conduction and synaptic delays in visual pathways~\cite{nowak1998axons,stepan2009delay}. More recent work shows that such delays support predictive, phase-advanced responses that compensate for processing lags in motion perception~\cite{nijhawan1994,priebe2006,hogendoorn2020}. Studies of motion processing in visual cortex and MT further highlight the importance of distributed and delay-based mechanisms~\cite{gundavarapu2019}.

Motivated by these findings, we emphasize the delayed coupling of two orientation rings as a simple but biologically grounded mechanism for distributed motion coding. The inclusion of delays in neural field models has a long history, beginning with Wilson–Cowan models~\cite{wilson1973mathematical} and subsequent analytical studies of wave propagation~\cite{pinto2001spatially,coombes2003waves}. More recently, the analysis of delay-coupled bumps in layered neural fields showed that axonal propagation can extend the timescale of short-term memory and make persistent activity more robust in the presence of fluctuations~\cite{kilpatrick2015delay} . These models establish a clear foundation for incorporating delay terms into neural fields, which we follow here. Here, we show that such a framework naturally generates illusion-like percepts through intrinsic delay-driven dynamics.”

In the following section, we introduce our delay-coupled neural field model. We begin by constructing stationary bump solutions and analyzing their linear stability using an interface reduction, which recasts the full neural field dynamics into a pair of coupled delay differential equations governing the motion of bump interfaces. We then examine input-free traveling bump solutions, showing how delays generate a family of metastable propagation speeds. Next, we analyze the response to persistent rotating inputs, demonstrating how bumps can entrain to the input speed before relaxing to a stable traveling state. We extend this analysis to stroboscopic forcing, revealing how periodic pulsed inputs can produce bumps that travel in the opposite direction of the stimulus, consistent with the wagon wheel illusion. Finally, we briefly extend our framework to spatially dependent delays, illustrating how distance-dependent transmission times likewise give rise to families of metastable traveling bumps.


\section{The Model}
\label{sec:TheModels}

Delays in neural processing are a fundamental feature of the visual system, arising from finite axonal conduction velocities, synaptic transmission times, and dendritic integration~\cite{stepan2009delay,nowak1998axons}. These delays are particularly relevant in motion perception, where visual cortex appears to compensate through predictive or phase-advanced activity~\cite{nijhawan1994,hogendoorn2020}. Motivated by these findings and prior delayed neural field studies~\cite{coombes2003waves,hutt2003,laing2006,kilpatrick2015delay}, we model two reciprocally coupled neural fields with delayed interlaminar communication, representing long-range interactions between visual populations. This delayed coupling captures the impact of axonal propagation times on spatiotemporal dynamics and motion-related perceptual artifacts.

\begin{figure}[t!]

        \centering
        \includegraphics[width=0.55\textwidth]{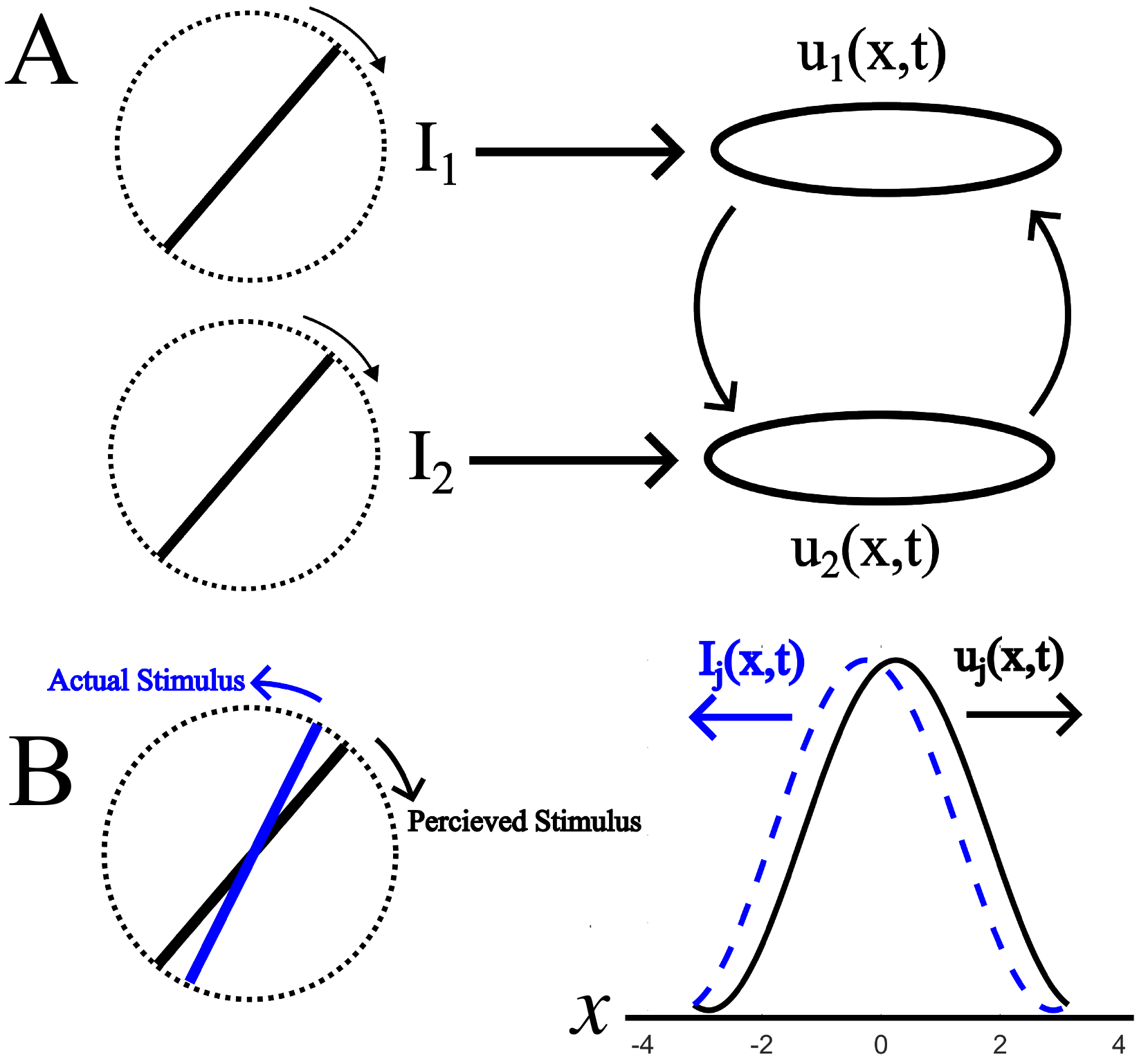}
        \vspace{-2mm}
        \caption{{\bf A.} Schematic of the delay-coupled ring model \eqref{dubdelay}. Two orientation rings receive rotating inputs $I_1$ and $I_2$ and evolve according to within-layer coupling $w(x)$ and delayed cross-layer coupling $w_c(x)$. This abstraction can be interpreted as orientation-selective populations in distinct visual areas (e.g., V1 and MT) exchanging delayed signals. {\bf B.} Conceptual illustration of the wagon wheel illusion: although a stimulus rotates in one direction, stroboscopic sampling can generate the percept of motion in the opposite direction. In our model, this corresponds to a traveling bump solution propagating counter to the input. We introduce this figure here to foreshadow the main phenomenon analyzed in later sections: how delayed coupling in neural fields provides a minimal circuit mechanism for generating illusory percepts.}
    \label{fig:model}
    \vspace{-5mm}
\end{figure}

We consider a pair of coupled neural fields on a periodic ring:
\begin{subequations} \label{dubdelay}
\begin{align}
\pd_t u_1(x,t) &= - u_1(x,t) + w(x)*f(u_1(x,t)) + w_c(x)*f(u_2(x,t - \tau))+I_1(x,t), \\
 \pd_t u_2(x,t) &= -u_2(x,t) + w(x)*f(u_2(x,t)) + w_c(x)*f(u_1(x,t-\tau))+I_2(x,t).
\end{align}
\end{subequations}
where $u_j(x,t)$ is the total synaptic input at location $x \in [-\pi,\pi)$ at time $t$. The convolutions $w*f(u) = \int_{- \pi}^{\pi} w(x-y)f(u(y,t))dy$ represent within layer ($u_j(x,t)$) and delayed cross-layer ($u_k(x,t-\tau)$, $k \neq j$) connectivity. This abstraction may be interpreted as two orientation columns in different visual areas (e.g., V1 and MT) exchanging delayed signals, or as parallel subnetworks within a single area~\cite{benyishai1995,movshon1996}. Our analysis of solutions and their stability works for general distance-dependent connectivity $w(x-y)$ and $w_c(x-y)$, but to demonstrate we employ the simple low-order lateral inhibitory kernels
\begin{equation} \label{w}
    w(x) = \cos(x) \hspace{1cm} \text{and } w_c(x) = \bar{w} \cos (x),
\end{equation}
so $\bar{w}$ parameterizes the strength of interlayer connectivity. Cosine kernels are the lowest-order Fourier approximation to Mexican-hat connectivity, capturing local excitation and long-range inhibition while remaining analytically tractable~\cite{ermentrout1998neural}.
Note $\tau$ represents the axonal propagation time between layers. While distributed delays can capture distance- or orientation-dependent transmission times within a cortical area~\cite{hutt2003,coombes2003waves}, here we focus on the dominant latency imposed by interareal communication (e.g.~V1-to-MT), which is well approximated by a discrete delay.
The firing rate function is typically taken to be a non-decreasing nonlinearity such as $f(u) = \left[ 1 + \exp (- \gamma (u-\theta))\right]^{-1}$ for some gain $\gamma>0$ and threshold $\theta > 0.$ Explicit formulae for solutions are obtainable in the high gain limit ($\gamma \to \infty$) whereby
\begin{equation} \label{firingrate}
    f(u) \to H(u-\theta) = \begin{cases}
        0, \ \ &u < \theta\\
        1, \ \ &u \geq \theta.
    \end{cases}
\end{equation}
Finally, $I_j(x,t)$ is an external input term to neural field $j$ representing a rotating stimulus in the visual field.

A visual representation of the model is given in Fig.~\ref{fig:model}, which both schematizes the architecture of the coupled rings and anticipates the main perceptual phenomenon we analyze. Panel A highlights how each ring receives its own input and communicates with the other via delayed cross-coupling, an abstraction of orientation-selective populations in different cortical areas. Panel B provides a conceptual link to perception, illustrating how such a circuit can, in principle, generate the wagon wheel illusion: a stimulus rotating in one direction may be perceived as rotating in the opposite direction. We include this figure here not as a result but as a roadmap for what follows, showing how the ingredients of our model naturally establish stroboscopic and illusory behaviors we examine in later sections.

\section{Input-free Solutions} 
Interface methods for Heaviside neural fields provide powerful low-dimensional reductions by tracking the motion of bump edges. This approach has been widely developed in delay-free settings to study pattern stability, front initiation, and bounded-domain effects~\cite{coombes2012interface,gokcce2017dynamics,faye2018threshold}. More recent work has extended interface dynamics to novel neural field architectures~\cite{cihak2022distinct}, but without explicit transmission delays. Prior studies have incorporated delays into neural field models using collective-coordinate or phase reductions, notably \cite{kilpatrick2015delay}, where delayed interareal coupling was shown to alter bump stability and stochastic wandering. 

Here, we take the next step by deriving \emph{interface equations with hard delays} for bump edges. To our knowledge, this is the first time delays have been built directly into the interface reduction itself, so that delay times appear explicitly in the edge dynamics. Deriving these input-free solutions and their stability not only highlights the novelty of delay-embedded interface methods, but also helps frame our subsequent analysis of how moving, stroboscopic inputs interact with delayed neural circuits. In this way, the baseline bump dynamics developed here form the foundation for understanding the perceptual illusions explored later in the paper.

\subsection{Stationary Bumps} 
\begin{figure}[t!]
    \centering
        \includegraphics[width = 0.85\textwidth]{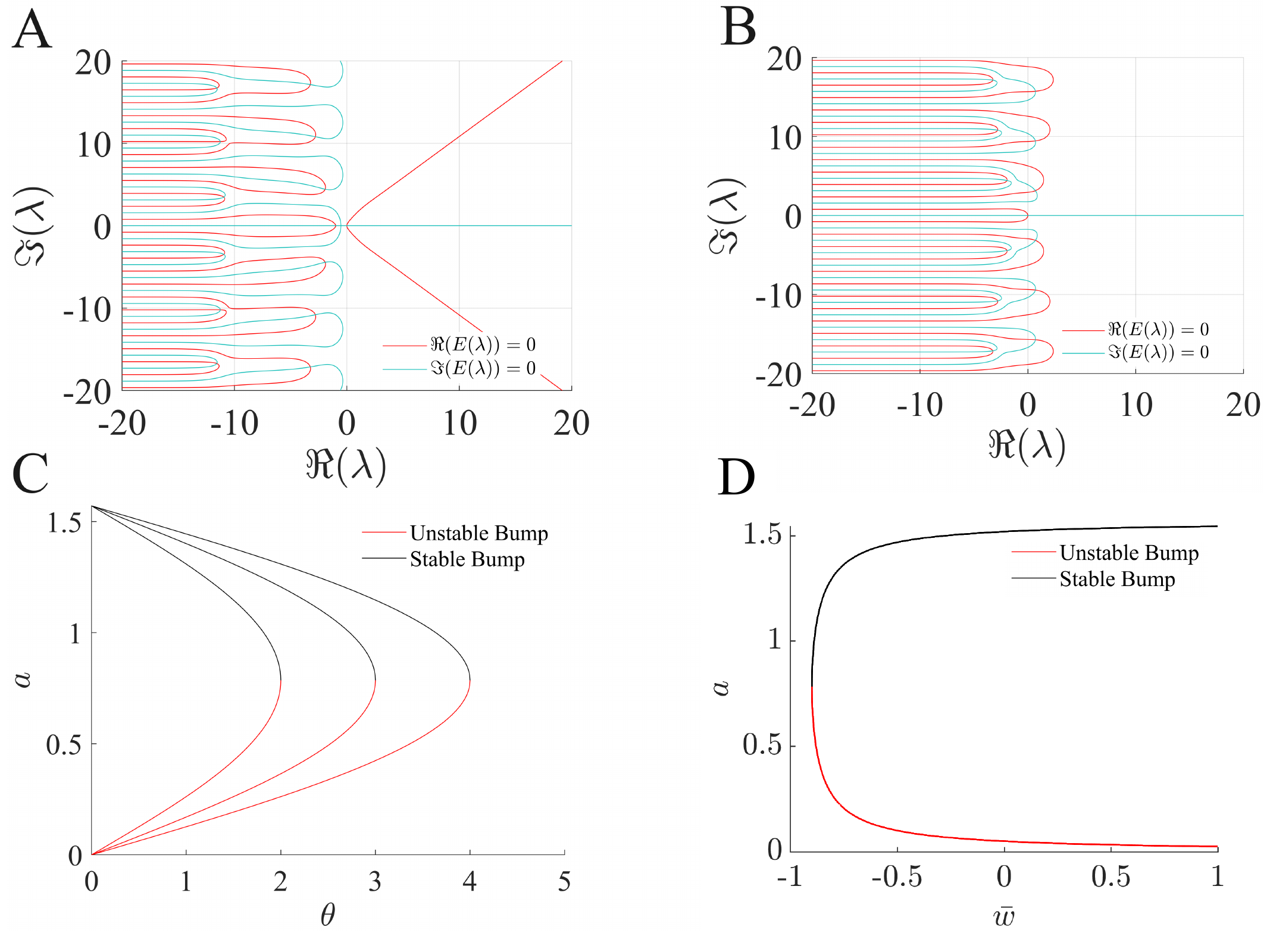}
        \vspace{-2mm}
        \caption{Half-width and stability of symmetric stationary bumps. {\bf A.} Eigenvalues determining stability of the wide bump occur at the zeros of the Evans function $E(\lambda)$ -- the intersection of the real and imaginary zero level sets. Here $\bar{w} = 1,$ $\tau = 10$ and $\theta = 0.1$ and we study the eigenvalues of the associate wide bump ($a_w$). {\bf B.}  Narrow bump ($a_n$) has a set of eigenvalues with positive real part.  {\bf C.} Stable/wide and unstable/narrow branches annihilate in a saddle node bifurcation as the threshold $\theta$ is increased to a critical value $\theta_c = 1 + \bar{w}$ for $\bar{w} = 1,2,3$.  {\bf D.} Saddle-node occurs at $\bar{w}_c = -1 + \theta$ as the cross-coupling strength is decreased ($\theta = 0.1$).} 
    \label{fig:stationary_stability}
    \vspace{-4mm}
\end{figure}

To orient our study of input-driven solutions, we begin by constructing bump solutions in the input-free case ($I(x,t) \equiv 0$). For stationary solutions, the delay $\tau$ is immaterial, since all delayed terms reduce to time-independent contributions. Let $u_j(x,t) = U_j(x).$ Then \eqref{dubdelay} reduces to
\begin{align*}
    U_j(x) = w(x)*f(U_j(x)) + w_c(x)*f(U_k(x)), \ \ j,k \in \{1,2\}, j \neq k.
\end{align*}
Even weight functions $w$ and $w_c$ admit even and unimodal single bumps $U_j(x)$ in each layer~\cite{amari1977}. For a Heaviside firing rate \eqref{firingrate}, we can identify bumps with their interface points $U(\pm a_j) = \theta$. We study co-located (i.e., isotopic or in-phase), but a similar analysis is possible when the two bump peaks are on opposite ends of the ring (allotopic as in \cite{folias2011new}). Co-located bumps have active regions ${\mc A}_j \equiv \{x \ | \ U_j (x)>\theta\}$ centered at 0 and satisfy
\begin{align} \label{interfaceeq}
\begin{split}
    U_j(x) &= \int_{-a_j}^{a_j}w(x-y)dy+\int_{-a_k}^{a_k}w_c(x-y)dy\\
    &=W(x+a_j)-W(x-a_j)+W_c(x+a_k)-W_c(x-a_k)
    \end{split}
\end{align}
where $W(x) = \int_{0}^xw(y)dy$ and $W_c(x) = \int_0^xw_c(y)dy$. Imposing self-consistency,
\begin{equation} \label{bumpthres}
    U_j(\pm a_j) = \theta = W(2a_j) + W_c(a_j+a_k) - W_c(a_j-a_k).
\end{equation}
Interface points $a_j$ may then be obtained by finding the roots of \eqref{bumpthres} and substituted into \eqref{interfaceeq} to find the bump profile. For instance given cosine connectivity \eqref{w}, symmetric bumps ($a_1 = a_2$) have a wide or narrow halfwidth
\begin{align*}
    a_w = \frac{\pi}{2}-\frac{1}{2}\arcsin{\left(\frac{\theta}{1+\bar{w}}\right)}, \ \ a_n = \frac{1}{2}\arcsin{\left(\frac{\theta}{1+\bar{w}}\right)}, \ \ \bar{w}+1 > \theta,
\end{align*}
where the wide/narrow bump is stable/unstable as in \cite{amari1977} and they annihilate in a saddle node bifurcation for $\bar{w} \to -1 + \theta$. We plot the dependence of $a$ on $\theta$ and $\bar{w}$ in  Fig.~\ref{fig:stationary_stability}. The profile is $U_j(x) = A\cos(x)$ with $A = 2(1+\bar{w})\sin(a)$ for $a \in \{a_n,a_w\}$.

We determine stability using the interface method of Amari \cite{amari1977}. Let $x_j^{\pm}(t)$ be the right ($+$) and left ($-$) interface points of $u_j(x,t)$, such that $u_j(x_j^{\pm}(t),t) = \theta$ for $j=1,2$. Differentiating with respect to $t$ and rearranging then gives
\begin{equation} \label{interface}
    \frac{dx_j^{\pm} (t)}{dt} = -\frac{\frac{\partial}{\partial t}u_j(x_j^{\pm}(t),t)}{\frac{\partial}{\partial x}u_j(x_j^{\pm} (t),t)}.
\end{equation}
Approximating the profile's spatial derivative with its stationary form at the interface,
$\partial_x u_j(x_j^{\pm}(t),t) \approx U_j'(\pm a_j)$, and defining the (positive) mobility
\[
\gamma_j \equiv \frac{1}{|U_j'(\pm a_j)|},
\]
we obtain for Heaviside firing rates
\begin{align}
\gamma_j^{-1} = w(0)-w(2a_j)+w_c(a_j-a_k)-w_c(a_j+a_k). \label{gamma}
\end{align}
Substituting in \eqref{gamma} and evaluating \eqref{dubdelay} at $x_j^{\pm}(t)$ we transform \eqref{interface} into
\begin{align*}
        &\frac{dx_j^{\pm}}{dt} = \mp \gamma_j\bigg[-\theta + W\big(x_j^+-x_j^-\big)+W_c\big(x_k^+(t-\tau)-x_j^{\pm} \big)-W_c\big(x_k^-(t-\tau)-x_j^{\pm} \big)\bigg].
\end{align*}
This highlights a key novelty of our approach: unlike previous interface analyses of delay-free neural fields~\cite{coombes2012interface,gokcce2017dynamics,faye2018threshold,cihak2022distinct}, our formulation produces bump-edge dynamics with \emph{explicit hard delays}, in which interface motion depends directly on past interface positions, thereby capturing axonal transmission times at the level of edge dynamics.
Changing variables $x_j^{\pm} = \pm a_j + z_j^{\pm}$ for $z_j^{\pm}$ small, we Taylor expand to first order,
\begin{align*}
        \frac{dz_j^{\pm}(t)}{dt} = &\mp \gamma_j\bigg[\left[z_j^+ -z_j^- \right]w(2a_j)\\&+\left[z_k^+(t-\tau)-z_j^{\pm} \right]w_c(a_j\mp a_k)-\left[z_k^-(t-\tau)-z_j^{\pm} \right]w_c(a_j \pm a_k)\bigg],
\end{align*}
so the vector $\mathbf{z}(t)\equiv [ z_1^-(t),  z_1^+(t), z_2^-(t),  z_2^+(t)
]^T$ satisfies the delayed linear system
\begin{align}
    \mathbf{z}'(t) = \mathbf{A}\mathbf{z}(t)+\mathbf{B}\mathbf{z}(t-\tau), \label{zsys}
\end{align}
whereby the perturbations of bump interfaces evolve through coupling to current~\cite{kilpatrick2013interareal} and past~\cite{kilpatrick2015delay} interface locations according to the current and delayed interaction terms
\begin{align*}
    \mathbf{A} &= \begin{bmatrix}
        \alpha_1 & \beta_1 & 0 & 0\\
        \beta_1 & \alpha_1 & 0 & 0\\
        0 & 0  & \alpha_2&\beta_2\\
        0 & 0 & \beta_2 &\alpha_2
    \end{bmatrix}, \hspace{1cm} \mathbf{B} = \begin{bmatrix}
        0 & 0 & \eta_1 & \nu_1\\
        0 & 0 & \nu_1 & \eta_1\\
        \eta_2 & \nu_2 & 0 & 0\\
        \nu_2 & \eta_2 & 0 & 0
    \end{bmatrix},
\end{align*}
where
$\alpha_j = \gamma_j[-w(2a_j)-w_c(a_1+a_2)+w_c(a_1-a_2)]$,
$\beta_j = \gamma_j w(2a_j)$,
$\eta_j = -\gamma_j w_c(a_1-a_2)$,
and $\nu_j = \gamma_j w_c(a_1+a_2)$.
 Eigensolutions to \eqref{zsys} of the form $\mathbf{z}(t) = \mathbf{v} e^{\lambda t}$ then have eigenvalues $\lambda$ that are roots of the Evans function
\begin{align*}
    E(\lambda) = \left|\mathbf{A}+e^{-\lambda\tau}\mathbf{B}-\lambda \mathbf{I}\right|.
\end{align*}
Complex roots are found at the intersections of the level sets $\mathrm{Re},E(\lambda)=0$ and $\mathrm{Im},E(\lambda)=0$. We numerically identify and plot these curves in Fig.~\ref{fig:stationary_stability} for cosine weights \eqref{w}, revealing that wide bumps are linearly stable while narrow bumps are unstable, consistent with classical results for neural fields~\cite{amari1977,ermentrout1998neural}. Unlike delay-free systems, where stability is governed by a finite-dimensional eigenvalue problem, the presence of transmission delays introduces an infinite spectrum through the transcendental dependence on $e^{-\lambda\tau}$. This structure produces repeated branches of eigenvalues and provides a natural mechanism for rich temporal dynamics, which will play a central role in the entrainment and stroboscopic forcing regimes analyzed in subsequent sections.

\subsection{Traveling Bumps}
Our model assumes that neural activity represents a percept of an object's instantaneous angular position, encoded by the peak of an activity profile on a ring. As this representation evolves in time, we are naturally led to study how rotating angles are represented by moving activity bumps. Canonical solutions of this type are traveling waves (or traveling bumps) that propagate around the ring with constant angular velocity $c$.

A key feature of the delay-coupled system studied here is that admissible traveling speeds are not continuously tunable, but instead form a discrete set determined by the interaction of recurrent coupling and transmission delays. We therefore seek traveling bump solutions of the form $u_j(x,t)=U_j(\xi)$, where $\xi=x-ct$ is the comoving coordinate associated with propagation speed $c$. Delayed activity then satisfies $u_j(x,t-\tau)=U_j(\xi+c\tau)$. In the comoving frame $(\xi,t)$, the governing equations reduce to a first-order system for the traveling-wave profiles,
\begin{align*}
- c\,U_j'(\xi) = - U_j(\xi) + (w * f(U_j))(\xi) + (w_c * f(U_k))(\xi + c\tau),
\end{align*}
where the convolution is now taken with respect to $\xi$. Assuming a Heaviside firing rate \eqref{firingrate} and exploiting translational symmetry, we may take $U_1(\xi)$ to be even about $\xi=0$, while $U_2(\xi)$ is even about $\xi=s$ (i.e., $U_2(\xi)=\tilde U_2(\xi-s)$ with $\tilde U_2$ even). As in the stationary case, traveling bumps possess interface points $a_j$ defined by the threshold conditions $U_1(\pm a_1)=\theta$ and $U_2(s\pm a_2)=\theta$.
Together, these assumptions reduce the problem to a pair of differential equations coupled through the bump interfaces:
\begin{align*}
        U_1'(\xi)  -\frac{1}{c}U_1(\xi) = -\frac{1}{c}\Gamma_1(\xi), \hspace{1cm}
        U_2'(\xi)  -\frac{1}{c}U_2(\xi) = -\frac{1}{c}\Gamma_2(\xi),
\end{align*}
where $\Gamma_j(\xi)$ is the sum of the within-layer and delayed cross-layer contributions:
\begin{align*}
        \Gamma_j(\xi) =& W(\xi-a_j - s_j)-W(\xi+a_j - s_j) \\
        &+ W_c(\xi-a_k+c\tau-s_k)-W_c(\xi+a_k+c\tau-s_k),
\end{align*}
where $s_1 = 0$ and $s_2 = s$. Using an integrating factor, we obtain
\begin{align*}
U_j(\xi)
&=e^{\xi/c}\left[\theta e^{-(s_j-a_j)/c}-\int_{s_j-a_j}^{\xi}\frac{e^{-z/c}}{c}\Gamma_j(z)\,dz\right].
\end{align*}
Imposing the threshold condition at the right interface then gives
\begin{align*}
\theta
&=U_j(s_j+a_j)
=e^{(s_j+a_j)/c}\left[\theta e^{-(s_j-a_j)/c}-\int_{s_j-a_j}^{s_j+a_j}\frac{e^{-z/c}}{c}\Gamma_j(z)\,dz\right].
\end{align*}

In general, self-consistency of the threshold equations $U_1(\pm a_1) = U_2(s \pm a_2) = \theta$ can be checked and numerically determined for arbitrary weight functions to identify the width and speed of traveling bumps. We generally expect to encounter a similar saddle-node bifurcation structure as in the stationary bump case. However, the analysis is most transparent with cosine weights \eqref{w}, where explicit solutions for traveling bumps can be derived and their stability properties characterized in detail. Substituting the ansatz
\begin{align*}
    U_1(\xi) = A \cos (\xi), \hspace{1cm} U_2(\xi) = B \cos (\xi -s),
\end{align*}
into the traveling wave equations then yields the system,
\begin{align*}
    A &= 2 \sin (a_1) + 2 \bar{w} \sin (a_2) \cos (s-c \tau), \\
    B &= 2 \sin (a_2) + 2 \bar{w} \sin (a_1) \cos (c \tau), \\
    cA &=  -2 \bar{w} \sin (a_2) \sin (s - c \tau), \\
    cB &= - 2 \bar{w} \sin (a_1) \sin (c \tau),
\end{align*}
and the threshold conditions enforce, noting that the second-layer bump is centered at $\xi=s$,
\begin{align*}
    U_1(a_1) = A \cos (a_1) = \theta, \hspace{1cm} 
    U_2(a_2) = B \cos (a_2 - s) = \theta.
\end{align*}
In what follows we focus on the co-located (isotopic) traveling-bump branch with $s=0$ for analytical tractability. 
This branch exists across the parameter regimes we study and, in our numerical simulations, is typically the stable attractor. 
Phase-offset states with $s \neq 0$ are not pursued here, as they introduce additional algebraic complexity and are deferred to future work.
Assuming $s=0$ and equal half-widths ($a_1=a_2=a$), self-consistency yields
\begin{align*}
    f(c) \equiv c + c \bar{w} \cos (c \tau) + \bar{w} \sin (c \tau) = 0.
\end{align*}
If $|\bar{w}| \geq 1$, then for any $n \in \mathbb{Z}$,
\begin{align*}
    f\left(\frac{2n\pi}{\tau}\right) \cdot f\left(\frac{(2n+1)\pi}{\tau}\right) \leq 0,
\end{align*}
indicating that $f(c)$ changes sign on each interval $\left(\frac{2n\pi}{\tau}, \frac{(2n+1)\pi}{\tau}\right)$. By the Intermediate Value Theorem, this guarantees the existence of at least one real root in each such interval, and thus $f$ admits infinitely many real roots.
Moreover, when $|\bar{w} | \geq 1$, a regular expansion shows the large-$|n|$ roots asymptotically satisfy
\begin{align*}
    c \sim \frac{2 \pi n}{\tau} \pm \frac{1}{\tau}\arccos\left(-\frac{1}{\bar{w}}\right) + \frac{1}{2\pi n }.
\end{align*}
Using $f(c)=0$, the amplitude reduces to $A = 2\sin(a)\left(1+\bar{w}\cos(c\tau)\right)$. Evaluating at the interface $\xi = a$ gives $\theta = A \cos(a) = \sin(2a)\left(1 + \bar{w} \cos(c\tau)\right)$, from which the two possible half-widths follow:
\begin{align}
    a_w = \frac{\pi}{2} - \frac{1}{2} \arcsin\left(\frac{\theta}{1 + \bar{w} \cos(c\tau)}\right), \hspace{1cm}  a_n = \frac{1}{2} \arcsin\left(\frac{\theta}{1 + \bar{w} \cos(c\tau)}\right), \label{twwid}
\end{align}
provided $|1+\bar{w}\cos(c\tau)|>\theta$.

\begin{figure}[t!]

    \begin{center} \includegraphics[width=13cm]{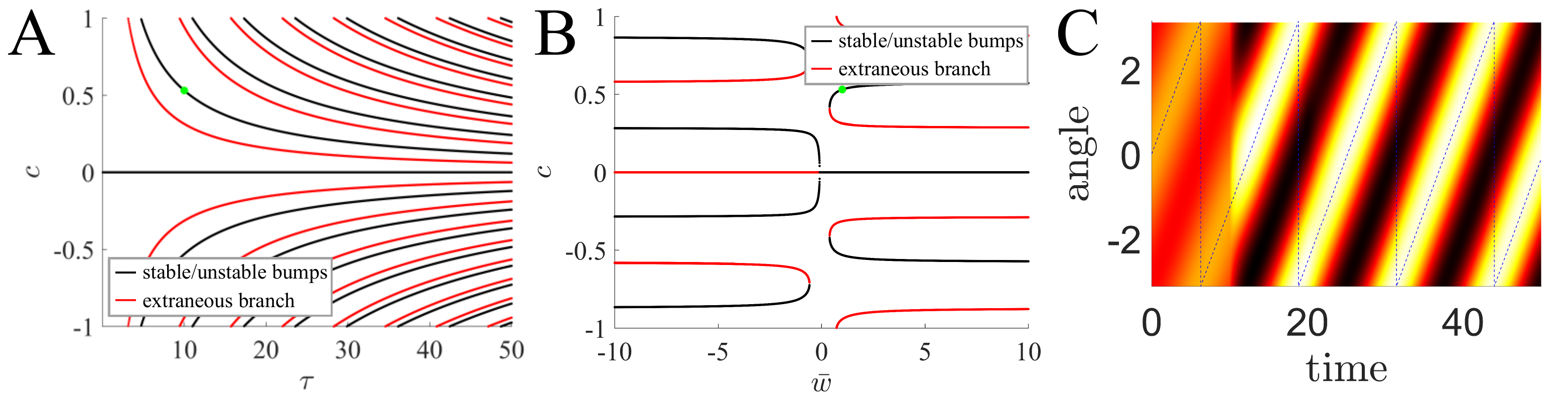} \end{center}
    \caption{\textbf{Families of metastable traveling bumps generated by delayed coupling.}
\textbf{(A)}~Traveling bump speeds $c$ as functions of the interlayer delay $\tau$ for fixed coupling strength $\bar{w}$. Black curves correspond to input-free traveling bump solutions of the delay-coupled neural field equations, each containing a wide stable and narrow unstable bump. Red curves indicate extraneous branches on which a proper bump width cannot be defined. Delays generate infinitely many branches of admissible propagation speeds. Here $\theta = 0.1$ and $\bar{w} = 1.$
\textbf{(B)}~Dependence of traveling bump speeds $c$ on the interlayer coupling strength $\bar{w}$ for fixed delay $\tau = 20$. Stable and unstable branches coexist and exchange stability at saddle-node bifurcations.
\textbf{(C)}~Spatiotemporal evolution of neural activity for a stable traveling bump corresponding to the green dot in \textbf{(A)} and \textbf{(B)}. Color indicates activity on the orientation ring as a function of time, with diagonal bands corresponding to coherent bump motion at a constant speed predicted by the interface reduction.}
    \label{fig:travelling_a}
\end{figure}

Figure~\ref{fig:travelling_a} summarizes the structure of traveling bump solutions generated by delayed interlayer coupling. Panels \textbf{A} and \textbf{B} show that axonal delays produce multiple discrete branches of admissible propagation speeds whose locations shift as the propagation delay $\tau$ and coupling strength $\bar w$ vary, yielding a lattice of coexisting traveling states over broad parameter ranges.
Black branches correspond to admissible solutions (for which (\ref{twwid}) yields a consistent bump pair), while red branches are extraneous (for which (\ref{twwid}) fails to produce a plausible pair of bumps).
When traveling bumps exist, wide (narrow) bumps are stable (unstable).
Panel \textbf{C} shows a representative stable traveling bump whose long-time speed agrees with the prediction of the interface reduction.

To analyze the linear stability of traveling bumps, we again apply the interface method, perturbing the left and right threshold crossings of each bump. 
The resulting four-dimensional system naturally admits a decomposition into translational and shape (width) perturbation modes of the traveling bump.
We assume co-located bumps in both layers with equal half-width $a$, such that $U_j(\pm a) = \theta$ for $j = 1,2$.
Denote the right and left interfaces as $x_j^\pm(t)$, and introduce small perturbations about the nominal traveling positions:
\[
x_j^\pm(t) = ct \pm a + z_j^\pm(t), \quad j = 1,2.
\]
The dynamics of the interfaces follow from the implicit definition $u_j(x_j^\pm(t),t) = \theta$. Differentiating both sides in time and applying the chain rule gives
\begin{equation}
\frac{dx_j^\pm(t)}{dt} = -\frac{\partial_t u_j(x_j^\pm(t),t)}{\partial_x u_j(x_j^\pm(t),t)}.
\label{eq:travel_interface}
\end{equation}
Differentiating gives $\frac{d}{dt}x_j^\pm(t) = c + \frac{d}{dt}z_j^\pm(t)$, and inserting into \eqref{eq:travel_interface} yields
\[
c + \frac{d}{dt}z_j^\pm(t) = -\frac{\partial_t u_j(ct \pm a + z_j^\pm(t),t)}{\partial_x u_j(ct \pm a + z_j^\pm(t),t)}.
\]
Expanding the right-hand side to first order in $z_j^\pm(t)$, and using the structure of the delayed bump equation, we derive linearized evolution equations for the four interface perturbations. Letting $\gamma_j = 1/|\partial_x U_j(\pm a)|$ denote the mobility (inverse slope magnitude) at threshold, and assuming symmetric bumps (so that $\gamma_1 = \gamma_2 = \gamma$), we find
\begin{align*}
\frac{dz_j^\pm(t)}{dt} &= \mp \gamma_j \Big[ \left(z_j^+(t) - z_j^-(t)\right) w(2a) + \left(z_k^+(t - \tau) - z_j^\pm(t)\right) w_c(a \mp a - c \tau) \\
&\quad - \left(z_k^-(t - \tau) - z_j^\pm(t)\right) w_c(a \pm a - c \tau) \Big],
\end{align*}
where \( j, k \in \{1,2\}, j \neq k \), and the arguments of the coupling weights reflect the influence of the delayed bump shape. Collecting the interface perturbations into a single vector,
\[
\mathbf{z}(t) = \begin{bmatrix} z_1^-(t) \\ z_1^+(t) \\ z_2^-(t) \\ z_2^+(t) \end{bmatrix},
\]
the linearized system takes the form
\begin{equation}
\frac{d\mathbf{z}}{dt} = \mathbf{A} \mathbf{z}(t) + \mathbf{B} \mathbf{z}(t - \tau),
\label{eq:travel_zsys}
\end{equation}
with interaction matrices
\begin{align*}
\mathbf{A} &= \begin{bmatrix}
\alpha & \beta & 0 & 0 \\
\beta & \alpha & 0 & 0 \\
0 & 0 & \alpha & \beta \\
0 & 0 & \beta & \alpha
\end{bmatrix}, \quad
\mathbf{B} = \begin{bmatrix}
0 & 0 & \eta & \nu \\
0 & 0 & \nu & \eta \\
\eta & \nu & 0 & 0 \\
\nu & \eta & 0 & 0
\end{bmatrix}, \label{eq:travel_AB}
\end{align*}
and coefficients
\begin{align*}
\alpha &= \gamma\Big[-w(2a)-w_c(2a-c\tau)+w_c(-c\tau)\Big], 
\qquad \beta = \gamma\, w(2a),\\
\eta &= -\gamma\, w_c(-c\tau), \qquad \nu = \gamma\, w_c(2a-c\tau).
\end{align*}

Linear stability of each traveling branch is determined by perturbations
to the bump interfaces, which naturally decompose into translational
and shape modes. Assuming exponential solutions of the form $\mathbf{z}(t) = \mathbf{v} e^{\lambda t}$, we obtain the characteristic equation
\begin{align*}
\det\left( \mathbf{A} + e^{-\lambda \tau} \mathbf{B} - \lambda \mathbf{I} \right) = 0,
\end{align*}
which defines the associated Evans function for traveling bumps. Its roots $\lambda$ determine linear stability: the traveling bump is linearly stable if and only if $\Re(\lambda)<0$ for all eigenvalues. As in the stationary case, stability is determined by the roots of the Evans function \eqref{eq:travel_zsys}, but here the coefficients depend explicitly on the traveling speed $c$, so stability varies across the family of admissible propagation speeds.

This four-interface reduction captures both translational and shape perturbations of traveling bumps and provides a direct extension of stationary interface methods to delay-coupled neural fields. Our results demonstrate that delay alone, even in the absence of external input, induces multistability in propagation speed, a feature that plays a central role in the entrainment and stroboscopic forcing mechanisms studied in subsequent sections.

\section{Entrainment by Continuous Motion}
\begin{figure*}[t!]
    \centering
    \includegraphics[width=0.85\textwidth]{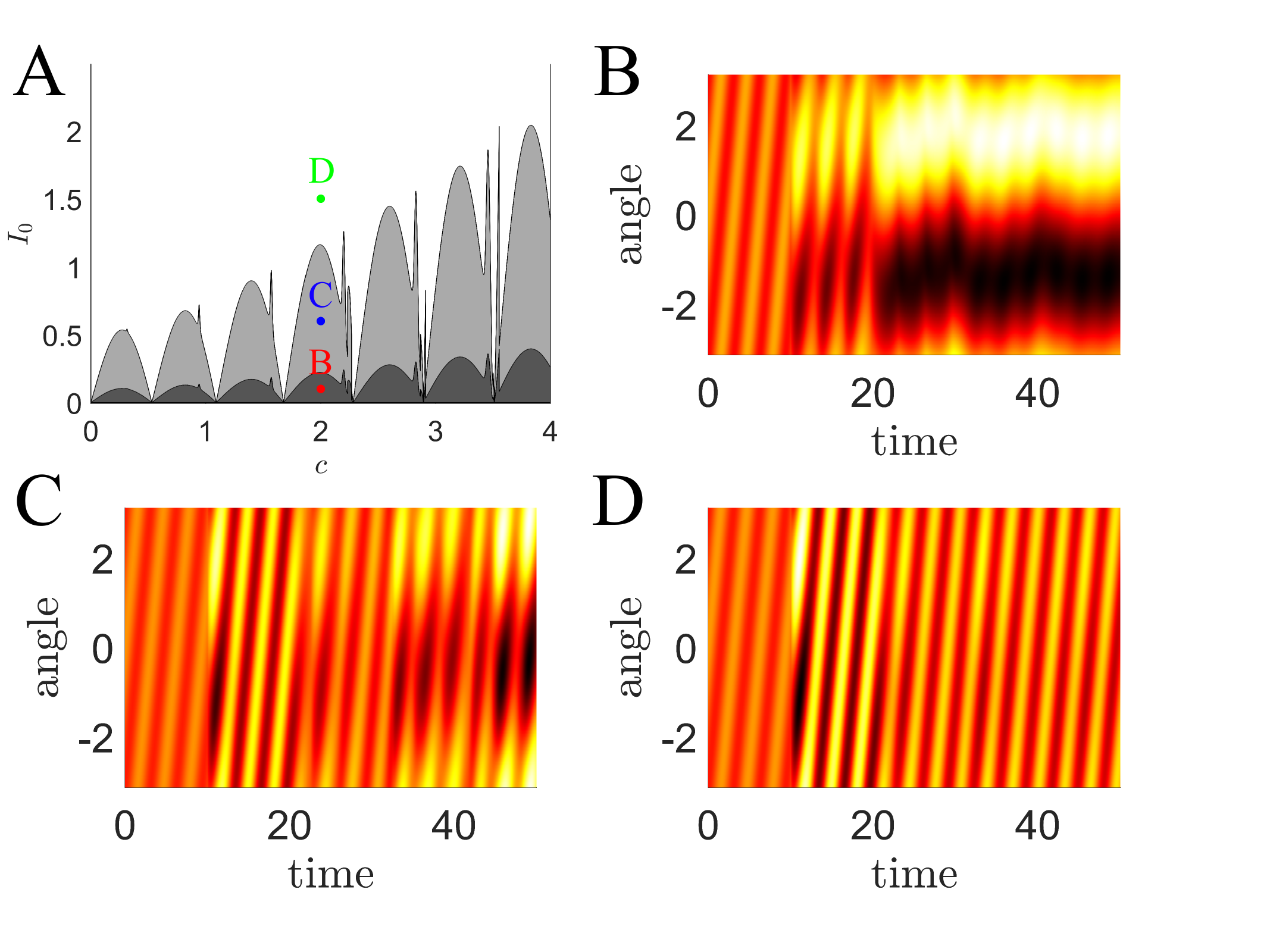}
    \caption{\textbf{Entrainment of traveling bumps to continuous motion.}
    \textbf{(A)} Regions of entrainment in the $(c,I_0)$ plane. Dark gray indicates parameter values for which no entrained traveling bump exists. Light gray indicates existence of an entrained solution that is linearly unstable. White indicates existence of a linearly stable entrained solution. Colored markers denote parameter values used in panels \textbf{(B)}--\textbf{(D)}.
    Panels \textbf{(B)}--\textbf{(D)} show the spatiotemporal evolution of the neural field $u_1(x,t)$ for the corresponding parameter values.
    \textbf{(B)} Below the entrainment threshold (red marker), showing failure of tracking and loss of coherent motion representation.
    \textbf{(C)} Unstable entrained solution (blue marker), illustrating transient tracking followed by drift or distortion of the bump profile.
    \textbf{(D)} Stable entrained solution (green marker), showing veridical motion perception: the bump remains coherent and rotates at the same speed and in the same direction as the stimulus, with a constant phase lag.
    Parameters: $\bar{w}=1$, $\theta=0.1$, $\tau=10$.}
    \label{entrainment}
\end{figure*}

Before turning to stroboscopic forcing, we first examine how delayed neural fields respond to continuously rotating stimuli.
Entrainment of a traveling bump to a moving input corresponds to veridical motion perception, meaning that the neural representation, encoded by the bump centroid, rotates at the same speed and in the same direction as the stimulus, without aliasing or reversal.
We work in the comoving coordinate $\xi=x-ct$ and consider inputs of the form
\[
I_j(x,t)=I(\xi).
\]
Seeking traveling profiles $u_j(x,t)=U_j(\xi)$ gives
\begin{align*}
-c\,U_j'(\xi)
= -U_j(\xi) + (w * f(U_j))(\xi) + (w_c * f(U_k))(\xi+c\tau) + I(\xi),
\end{align*}
where $(w*g)(\xi)=\int_{-\pi}^{\pi}w(\xi-y)g(y)\,dy$.
Letting $s_j\pm a_j$ denote the threshold crossings of $U_j$ (so $U_j(s_j\pm a_j)=\theta$) yields
\begin{align*}
U_j'(\xi)-\frac{1}{c}U_j(\xi)
= -\frac{1}{c}\Lambda_j(\xi)-\frac{1}{c}I(\xi).
\end{align*}
where 
\begin{align*}
    \Lambda_j(\xi) =& W(\xi-s_j+a_j)-W(\xi-s_j-a_j) \\
    &+W_c(\xi+c\tau-s_k+a_k)-W_c(\xi+c\tau-s_k-a_k).
\end{align*}
Here $W$ and $W_c$ denote antiderivatives of the within- and cross-layer kernels as before.
Integrating with the integrating factor $e^{-\xi/c}$ and applying the reference threshold
condition $U_j(s_j-a_j)=\theta$ yields
\begin{align*}
U_j(\xi)
= e^{\xi/c}\bigg[\theta e^{-(s_j-a_j)/c}
-\int_{s_j-a_j}^{\xi}\frac{e^{-z/c}}{c}\Lambda_j(z)\,dz
-\int_{s_j-a_j}^{\xi}\frac{e^{-z/c}}{c}I(z)\,dz\bigg].
\end{align*}
For cosine weight kernels and $I(\xi)=I_0\cos(\xi)$, the traveling bump profile can be written explicitly as
\begin{align*}
U_j(\xi)
=&\frac{-\big(A_j-I_0\cos(s_j)\big)-\big(B_j+I_0\sin(s_j)\big)c}{c^2+1}\,\cos(\xi-s_j)
 \\
 & + \frac{\big(A_j-I_0\cos(s_j)\big)c-\big(B_j+I_0\sin(s_j)\big)}{c^2+1}\,\sin(\xi-s_j) \\
&+ \frac{\cos(a_j)\,e^{(\xi+a_j-s_j)/c}}{c^2+1}
\Big[\big(A_j-I_0\cos(s_j)\big)+\big(B_j+I_0\sin(s_j)\big)c\Big] \\
&+ \frac{\sin(a_j)\,e^{(\xi+a_j-s_j)/c}}{c^2+1}
\Big[\big(A_j-I_0\cos(s_j)\big)c-\big(B_j+I_0\sin(s_j)\big)\Big]
 \\
 & + \theta\,e^{(\xi+a_j-s_j)/c},
\end{align*}
where 
\begin{align}
\begin{split}
    A_j &= -2\sin(a_j)-2\bar{w}\sin(a_k)\cos(c\tau+s_j-s_k)\\
    B_j &= 2 \bar{w} \sin(a_k)\sin(c\tau+s_j-s_k).
\end{split} \label{ABdef}
\end{align}
Imposing self-consistency for a shifted-even bump, reflecting about $\xi=s_j$, removes the $\sin(\xi-s_j)$ component, and enforcing periodicity removes the nonperiodic homogeneous contributions.
This yields the algebraic entrainment conditions
\begin{align}
    \begin{split}
        (A_j-I_0\cos(s_j))c-(B_j+I_0\sin(s_j)) &= 0\\
        \frac{\big[(A_j-I_0\cos(s_j))+(B_j+I_0\sin(s_j))c\big]\cos(a_j)}{c^2+1}+\theta &=0.
    \end{split} \label{travelling_self_con}
\end{align}
If these four equations have solutions, then an entrained solution exists that tracks the input, potentially with a phase lag.
The first condition enforces cancellation of the sine component in the profile, while the second enforces threshold crossing at the bump interface.
Equations \eqref{travelling_self_con} thus define an entrainment (phase-locking) manifold in $(a_j,s_j)$ for fixed $(c,\tau,\bar{w},I_0)$, corresponding to traveling bump solutions locked to the stimulus in the rotating frame.
While the system \eqref{travelling_self_con} may have some solutions that permit bumps of different profiles and relative shifts, such solutions may only be investigated numerically. For the sake of analytical tractability, assume $a_1 = a_2 = a$ and $s_1 = s_2 = s,$ i.e. both solutions share the same profile with the same shift. In this case, we may explicitly solve for the input amplitude:
\begin{equation}
    I_0 = \bigg[\left[f(a)-cg(a)\right]^2+g^2(a)\bigg]^{1/2}
\end{equation} where
\begin{align*}
        f(x) &= -2\left[c+c\bar{w}\cos(c\tau)+\bar{w}\sin(c\tau)\right]\sin(x),\\
        g(x) &= \theta \sec(x)-2\left[1+\bar{w}\cos(c\tau)\right]\sin(x).
\end{align*}
Hence for a given speed, we minimize over $a \in [0,\pi]$ to find the smallest input amplitude that allows for entrainment:
\begin{equation}
    I_{\rm min}(c) = \min_{a \in [0,\pi]}\bigg[\left[f(a)-cg(a)\right]^2+g^2(a)\bigg]^{1/2}.
\end{equation}
In Fig.~\ref{entrainment}, we plot the regions in the $(c,I_0)$ plane that admit entrainment of a traveling bump to a continuously rotating stimulus.
The fine, tooth-like structure of the entrainment boundary is controlled by the interlayer coupling strength $\bar{w}$ and the delay $\tau$, and arises from constructive and destructive interference between the moving input and delayed feedback from the opposing ring.
These narrow features correspond to resonant speed windows in which delayed coupling either enhances or suppresses phase locking, producing alternating bands of stability and instability.
At larger scales, the entrainment boundary is dominated by the firing threshold $\theta$, which sets the minimum input strength required to sustain a coherent bump translating at speed $c$.
As a result, sufficiently fast stimuli require proportionally stronger input to overcome the intrinsic tendency of the bump to decay, yielding the asymptotic scaling $I_{\min}(c)\sim \theta c$ for large $c$.

Since $\bigg[\left[f(a)-cg(a)\right]^2+g^2(a)\bigg]^{1/2}$ is unbounded for $a \in [0,\pi],$ an entrained solution exists for any $I_0 \geq I_{\rm min}(c)$ for any fixed $c$. The width of the solution is given by solving
\begin{equation*}
    I_0 = \bigg[\left[f(a)-cg(a)\right]^2+g^2(a)\bigg]^{1/2}
\end{equation*} for $a$ numerically. The resulting shift $s$ is given by
\begin{equation}
    s = -\sign(c)\left| \arccos\left(\frac{1}{I_0}g(a)\right)\right|.
\end{equation}
The sign is chosen so that the bump peak trails the direction of motion of the
input, consistent with a lagging entrained percept.

To allow layer-dependent forcing, we write the input amplitudes as $\bar I_j$ (the symmetric case corresponds to $\bar I_1=\bar I_2=I_0$) and briefly re-derive the traveling-wave entrainment conditions using a Fourier representation. This provides an independent consistency check on the interface analysis and allows for asymmetric forcing between layers. Writing the traveling-wave equation in the comoving coordinate $\xi=x-ct$,
\[
-cU_j'(\xi) = -U_j(\xi) + w*f(U_j) + w_c*f(U_k(\xi+c\tau)) + I_j(\xi),
\]
and expanding in Fourier modes
\[
\hat h(n)=\frac{1}{2\pi}\int_{-\pi}^\pi h(\xi)e^{-in\xi}\,d\xi,
\]
gives
\begin{equation}
-icn\hat U_j(n) = -\hat U_j(n) +2\pi\hat w(n)\mathcal F_n(f(U_j)) +2\pi\hat w_c(n)\mathcal F_n(f(U_k(\xi+c\tau))) +\hat I_j(n).
\end{equation}

Specializing to cosine connectivity and input, $w(x)=\cos x$, $w_c(x)=\bar w\cos x$, and $I_j(x)=\bar I_j\cos x$, only the modes $n=\pm1$ are nonzero. Assuming Heaviside firing rates and shifted even bump solutions with interfaces $s_j\pm a_j$, the nonvanishing Fourier coefficients satisfy
\begin{equation}
\hat U_j(\pm1)(1\mp ic) = e^{\mp is_j}\sin a_j +\bar w e^{\mp i(s_k-c\tau)}\sin a_k +\tfrac12\bar I_j .
\label{eq:fourier_coeff_reduced}
\end{equation}

Writing $\hat U_j(\pm1)=\tilde U_j e^{\mp is_j}$ and eliminating $\tilde U_j$ yields algebraic relations equivalent to the interface-based self-consistency conditions \eqref{travelling_self_con}. In particular, the amplitude–speed relation
\begin{equation}
(\bar I_j\cos s_j - A_j)c = -B_j - \bar I_j\sin s_j
\end{equation}
and the threshold condition
\begin{equation}
(A_j-\bar I_j\cos s_j)\cos a_j + \theta = 0
\end{equation}
are recovered exactly.

For asymmetric forcing, taking $\bar I_1=\bar I$ and $\bar I_2=0$ and defining the relative phase shift $r=s_1-s_2$, the traveling-wave solutions satisfy a closed system of four nonlinear algebraic equations for $(a_1,a_2,s_1,r)$, which can be solved numerically for a given speed $c$ and input amplitude $\bar I$.

We analyze the linear stability of co-located traveling bump solutions entrained to an input of the form $I_j(x,t) = I_0 \cos(x - ct)$. Assuming co-located bumps of equal width, the solution takes the form $U_j(\xi)$ with $\xi = x - ct$ and interface locations $x_j^\pm(t) = ct \pm a + z_j^\pm(t)$, where $z_j^\pm(t)$ denote small perturbations.

Differentiating $u_j(x_j^\pm(t),t)=\theta$ gives
\begin{align*}
\frac{dx_j^\pm}{dt}= -\frac{\partial_t u_j(x_j^\pm(t),t)}{\partial_x u_j(x_j^\pm(t),t)}.
\end{align*}
Using $x_j^\pm(t)=ct\pm a+z_j^\pm(t)$ (so $\dot x_j^\pm=c+\dot z_j^\pm$) yields an
evolution equation for $z_j^\pm$ after linearization.
Evaluating $\partial_t u_j$ from the full neural field and expanding to leading order gives
\begin{equation}
\frac{d z_j^\pm(t)}{dt} = \mp \gamma \left[ \Delta_j(t) \right] 
\end{equation}
where $\gamma = 1/|\partial_x U_j(\pm a)|$ is the inverse slope at threshold, and $\Delta_j(t)$ contains terms from intra-layer and inter-layer contributions, as in the input-free case.

Because the input is comoving, $I(x,t)=I(\xi)$ contributes only to the base
traveling solution and does not generate additional linear terms in the
interface equations.
Its influence on stability therefore enters only indirectly, through
modifications of the entrained bump width $a$ and the threshold slope
$\gamma$, which in turn alter the interaction matrices $\mathbf{A}$ and
$\mathbf{B}$.
This reflects the fact that continuously moving stimuli act as a smooth
reference frame for the neural representation, supporting veridical motion
tracking without introducing new dynamical instabilities.
In the next section, we show that this structure breaks down when motion is
sampled discretely in time.

\section{Dynamics and Discrete-Time Motion under Stroboscopic Forcing}
\label{sect:strobconst}
We now turn to stroboscopic forcing, in which the stimulus is sampled discretely in time.
Unlike continuous motion, stroboscopic inputs do not define a smooth comoving reference
frame for the neural activity. Instead, the bump evolves autonomously between stimulus
events and is intermittently corrected by brief input pulses.
The interaction between these discrete corrections and delayed interlayer coupling
fundamentally alters the dynamics, leading to multistability, speed quantization,
and strong sensitivity to input timing.
In the following, we first derive reduced dynamics for general stroboscopic inputs,
and then specialize to impulsive forcing, where the discrete-time structure of the
motion becomes explicit.

\subsection{General stroboscopic inputs}
Accordingly, we repeat the traveling-bump analysis under a temporally modulated input,
assuming the stimulus takes the separable form
\begin{equation*}
    I(\xi,t) = I_t(t)I_\xi(\xi).
\end{equation*}
Since the input has time-varying amplitude, we may no longer assume a solution traveling at a constant speed; instead, the bump position $p(t)$ evolves dynamically under the combined influence of delayed coupling and intermittent forcing.
Hence, we define the position of the bump to be $p(t)$ with $p$ differentiable, and additionally define $d(t) \equiv p(t)-ct.$
For analytical tractability, we assume identical solutions on both rings, so that the two layers remain phase-aligned and share a common bump position $p(t)$ and profile.
We therefore write
\[
u_j(x,t)=U(x-p(t)) = U(\xi-d(t)),
\]
which reduces the system to a single effective bump with time-dependent position.
Under these assumptions, the model reduces to
\begin{align}
\begin{split}
    -p'(t)U(\xi-d(t)) = &-U(\xi-d(t)) + w(\xi)*f(U(\xi-d(t)))\\
    &+ w_c(\xi)*f(U(\xi-d(t-\tau)+c\tau))+I_1(t)I_\xi(\xi).
    \end{split}
\end{align}
Once again assuming cosine connectivity functions and $\xi$ component of the input and interface points $\pm a,$ the equations for the Fourier modes of the solution are
\begin{equation} 
    (1\mp ip'(t))\hat{U}(\pm1,t) =e^{\mp id(t)}\sin(a)+\bar{w} e^{\mp i(d(t-\tau)-c\tau)}\sin(a)+\frac{1}{2}I_t(t). \label{stroboshiftcoeff}
\end{equation}
Then letting $\hat{U}(\pm 1,t) = \tilde{U}(t)e^{\mp id(t)},$ adding equations yields
\begin{equation}
    \tilde{U}(t) = \sin(a)+\bar{w}\cos[p(t)-p(t-\tau)]\sin(a)+\frac{1}{2}I_t(t)\cos[p(t)-ct] \label{strobocoeff}
\end{equation} and hence
\begin{equation}
    U(x,t) = \bigg[2\sin(a)+2\bar{w}\cos[p(t)-p(t-\tau)]\sin(a)+I_t(t)\cos[p(t)-ct]\bigg]\cos[x-p(t)]. \label{profileeq}
\end{equation}
Additionally, plugging \eqref{strobocoeff} into \eqref{stroboshiftcoeff} gives
\begin{align}
\begin{split}
    &-2\bar{w}\sin[p(t)-p(t-\tau)]\sin(a)-I_t(t)\sin[p(t)-ct] \\= &\bigg[2\sin(a)+2\bar{w}\cos[p(t)-p(t-\tau)]\sin(a)+I_t(t)\cos[p(t)-ct]\bigg]p'(t). \label{delayode1}
    \end{split}
\end{align}
Finally, evaluating \eqref{profileeq} at $x = p(t)+a$ and combining with \eqref{delayode1} yields the delay differential equation
\begin{equation}
    p'(t) = \frac{1}{\theta}\bigg(-2\bar{w}\sin[p(t)-p(t-\tau)]\sin(a)-I_t(t)\sin[p(t)-ct]\bigg)\cos(a). \label{p_delay}
\end{equation}
Equation \eqref{p_delay} describes the instantaneous velocity of the bump as the sum of a delayed coupling term, which favors intrinsic wave speeds of the input-free system, and a time-dependent forcing term that pulls the bump toward the current stimulus location.

To isolate the dynamical consequences of discrete sampling, we now specialize to impulsive forcing, which captures the essential features of stroboscopic motion while permitting analytical progress.
In this limit, the continuous-time bump dynamics reduce to a hybrid system with smooth delay-driven evolution punctuated by discrete phase resets.

\subsection{Impulse forcing and speed quantization}
We now consider impulsive forcing, taking $I_t(t)$ to be a delta train,
\[
I_t(t) = \sum_{n=0}^{\infty} A\,\delta(t-t_n),
\]
so that the bump evolves freely between impulses and undergoes instantaneous phase resets at the pulse times $t_n$.
This case is characterized by periods of input-free evolution, interrupted by instantaneous jumps in the bump position.
Each impulse at time $t=t_n$ produces a discontinuous change
\[
p(t_n^+)-p(t_n^-)= -\frac{A}{\theta}\sin[p(t_n)-ct_n]\cos(a).
\]
The bump dynamics then evolve autonomously, reducing to the unforced delay-coupled neural field with a discrete family of stable traveling speeds inherited from the input-free model.
The impulses act as instantaneous phase resets that can shift the solution between the basins of attraction of these speeds.
As we show below, these phase resets organize the long-term dynamics into \emph{quantized} traveling speeds, with transitions controlled by impulse amplitude and timing.
Consider a time interval with no input. In this case, \eqref{p_delay} reduces to 
$$p'(t) = -\frac{\bar{w}}{\theta}\sin[p(t)-p(t-\tau)]\sin(2a)$$
and evaluating \eqref{profileeq} at $x = p(t)+a$ yields
$$\bigg(1+\bar{w}\cos[p(t)-p(t-\tau)]\bigg)\sin(2a) = \theta.$$ Combining the two yields
\begin{equation}
    p'(t) = -\frac{\bar{w}\sin[p(t)-p(t-\tau)]}{1+\bar{w}\cos[p(t)-p(t-\tau)]}. \label{simplifiedp}
\end{equation}
We now consider inter-impulse solutions that translate at an approximately constant speed $r$,
corresponding to phase-locked solutions of the autonomous delay equation.
To analyze their stability and phase shifts, it is convenient to introduce the comoving phase variable
\[
\phi(t) = p(t)-rt,
\]
for which uniformly translating solutions correspond to constant $\phi(t)$.
Here $r$ denotes a candidate traveling speed selected by the autonomous dynamics.
The resulting autonomous phase equation becomes
\begin{equation}
    \phi'(t) = -\frac{\bar{w}\sin[\phi(t)-\phi(t-\tau)+r\tau]}{1+\bar{w}\cos[\phi(t)-\phi(t-\tau)+r\tau]}-r. \label{simplifiedphi}
\end{equation}
Now assume $\phi(t-\tau) \equiv s_1,$ and look for solutions $\phi(t) \equiv s_2.$
Defining $s = s_2-s_1$ and plugging in:
$$r+r\bar{w}\cos(s+r\tau)+\bar{w}\sin(s+r\tau) = 0$$ with solutions
\begin{align}
\begin{split}
    &s_a = -\arcsin\left(\frac{r}{\bar{w}\sqrt{r^2+1}}\right)-r\tau-\arctan(r) + 2\pi n,\\ &s_b=\pi+\arcsin\left(\frac{r}{\bar{w}\sqrt{r^2+1}}\right)-r\tau-\arctan(r) + 2\pi n, \label{entrainedsols}
    \end{split}
\end{align}
for $n\in \mathbb{Z}$. The delayed phase dynamics therefore admit multiple fixed points, separated by singular phase values that act as separatrices between distinct traveling-speed states. The right-hand side of \eqref{simplifiedphi} is undefined at each $s_b$, so these singular values cannot be crossed by continuous trajectories. Between successive singularities the phase dynamics are smooth and monotone, implying that each regular fixed point $s_a$ is locally attracting.

Consequently, the autonomous delay dynamics partition phase space into alternating stable branches $s_a$, separated by singular barriers $s_b$ that delineate their basins of attraction. In the special case where $r$ coincides with a natural stable speed of the input-free system, the stable phases satisfy $s_a=2\pi n$, so that the traveling wave remains unshifted modulo $2\pi$. More generally, impulsive perturbations that drive the phase across a separatrix $s_b$ can induce transitions between distinct stable traveling-speed states.

\begin{figure*}[t!]
     \centering
        \includegraphics[width=0.85\textwidth]{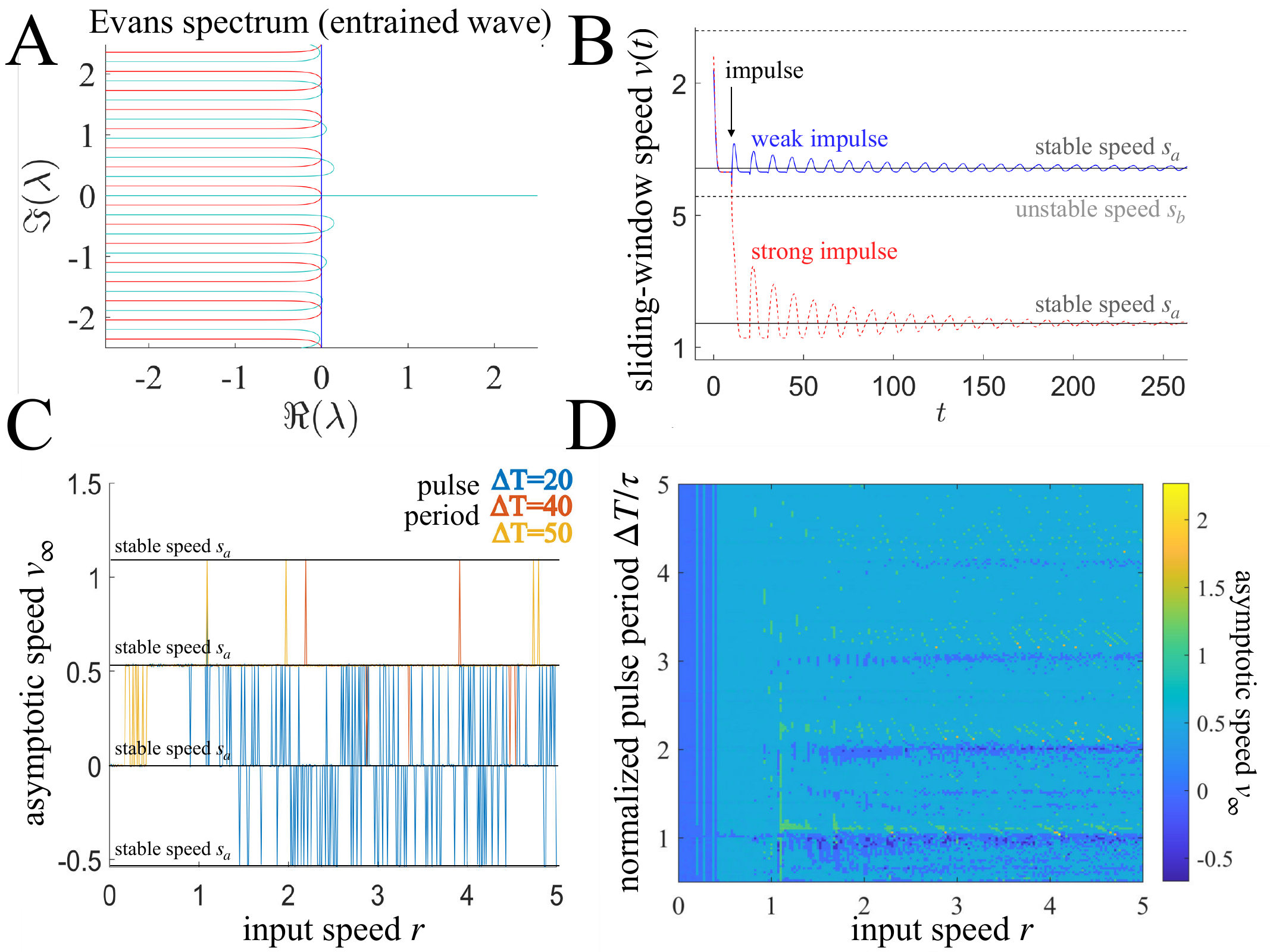}
    \caption{
    \textbf{Speed selection under impulsive forcing.}
    \textbf{(A)}~Contour plot of the Evans function \eqref{eq:evans} associated with the autonomous inter-impulse dynamics, illustrating the stability of discrete traveling speeds. Here $r =2$ and $\bar{w} =1.$
    \textbf{(B)}~Time series of the sliding-window speed $v(t)$ following a single impulse. A weak impulse (blue) does not cross a separatrix and relaxes back to the original speed, while a stronger impulse (red) induces a transition to a neighboring stable speed. The parameters are the same as panel A, with the impulse amplitudes at $0.5$ and $2$ respectively.
    \textbf{(C)}~Asymptotic speed $\lim_{t\to\infty} v(t)$ under repeated impulses applied at fixed intervals $\Delta T$, showing convergence to a discrete set of quantized speeds. Positive asymptotic speeds slower than the stimulus correspond to \emph{aliasing}, while negative speeds correspond to \emph{anti-aliasing} (motion reversal). The impulse amplitude $A$ is set at $1,$ with $\tau = 10,$ $\bar{w} = 1,$ and $\theta = 0.1.$
    \textbf{(D)}~Asymptotic speed as a function of input speed $r$ and pulse period normalized by the delay, $\Delta T/\tau$, revealing banded regions of aliasing ($0<v<r$) and anti-aliasing ($v<0$) organized by impulse timing. Parameters $\theta = 0.1$, $\bar{w} = 1$; impulse amplitude $A=1$; period $\Delta T$, and input speed $r$ are varied.}
    \label{fig:stroboscopic}
\end{figure*}

Input-free dynamics are governed by the autonomous delay equation \eqref{simplifiedphi}, whose phase-locked solutions determine the admissible traveling speeds of the bump.
To confirm that these phase-locked states persist between impulses, we briefly assess their linear stability.
Returning to \eqref{simplifiedphi} and assuming the ansatz 
$$\phi(t) = \varepsilon e^{\lambda t}$$
and taking a first order Taylor expansion, we find the eigenvalues $\lambda$ are roots of 
\begin{equation}
    E(\lambda)=\left(\frac{\bar{w}(\bar{w}+\cos(r\tau))}{(1+\bar{w}\cos(r\tau))^2}\right)\left(e^{-\lambda\tau}-1\right)-\lambda. \label{eq:evans}
\end{equation}
The Evans function \eqref{eq:evans} characterizes the linear stability of these phase-locked states of the inter-impulse dynamics, confirming that the admissible speeds persist as attracting solutions under small perturbations. An example visualization of these roots is given in Fig.~\ref{fig:stroboscopic}{\bf A}.

When $r$ coincides with a stable input-free speed, all eigenvalues have non-positive real part. We now consider the quantity
$$v(t) = \frac{1}{\tau}\left(p(t)-p(t-\tau)\right) = \frac{1}{\tau}\int_{t-\tau}^t p'(s)ds,$$ the sliding-window average speed of the bump. Differentiating the previous equation, we derive the following DDE for $v:$
\begin{equation}
    \tau v'(t) = -\frac{\bar{w}\sin(\tau v(t))}{1+\bar{w}\cos(\tau v(t))} +  \frac{\bar{w}\sin(\tau v(t-\tau))}{1+\bar{w}\cos(\tau v(t-\tau))}.
\end{equation}
Note this equation has the same linearization as \eqref{simplifiedphi} at the input-free stable speeds, and hence these speeds are also stable fixed points for $v(t).$ Additionally, this equation is not defined at the extraneous branches, and thus these speeds form barriers that $v(t)$ cannot normally cross. The only exception is when $v(t)$ is discontinuous, as is the case at $v(\tau^+).$ Since $p(t)$ is continuous after $t = 0^+,$ $v(t)$ is continuous after $t = \tau^+$ and hence remains trapped between whatever extraneous branches it finds itself between, becoming attracted to whatever stable speed is in the same region. Furthermore, we can easily derive a simple yet accurate approximation for $v(\tau^+).$ If $\tau$ is large enough, then by $t = \tau,$ $p'(t)$ will have almost relaxed to $r,$ whatever stable speed it was traveling at (or close to). Then, the average speed of $p$ from $0^+$ to $\tau^+$ will be approximately $r - \frac{\overline{A}}{\tau}$ where $\overline{A}$ is the amplitude of the delta pulse, modded between two consecutive $s_b$ such that $0$ lies between them. Whatever basin of attraction $r - \frac{\overline{A}}{\tau}$ lays in will be the stable speed the bump will tend to.

Impulsive inputs interact with delayed neural dynamics in a highly structured way, reshaping motion representations by selectively promoting or suppressing intrinsic traveling states. Figure~\ref{fig:stroboscopic} summarizes how impulsive forcing interacts with the delay-induced speed lattice. In the absence of input, the autonomous dynamics admit a discrete set of admissible traveling speeds, corresponding to linearly stable roots of the Evans function~\eqref{eq:evans} (Fig.~\ref{fig:stroboscopic}A).
A single impulse perturbs the system within this phase-space structure: weak perturbations remain within a basin of attraction and decay back to the same intrinsic speed, whereas sufficiently strong impulses can cross a separatrix and induce transitions to neighboring speed branches (Fig.~\ref{fig:stroboscopic}B). Under repeated impulsive forcing, these phase resets prevent full relaxation and instead drive convergence toward one of the admissible intrinsic speeds, producing quantized long-term motion (Fig.~\ref{fig:stroboscopic}C). The resulting selected speed depends jointly on the stimulus speed $r$ and the inter-impulse period $\Delta T$, giving rise to banded regions of speed selection that reflect the underlying delay-induced multistability (Fig.~\ref{fig:stroboscopic}D).

\begin{figure}[t!]
\begin{center} \includegraphics[width=\textwidth]{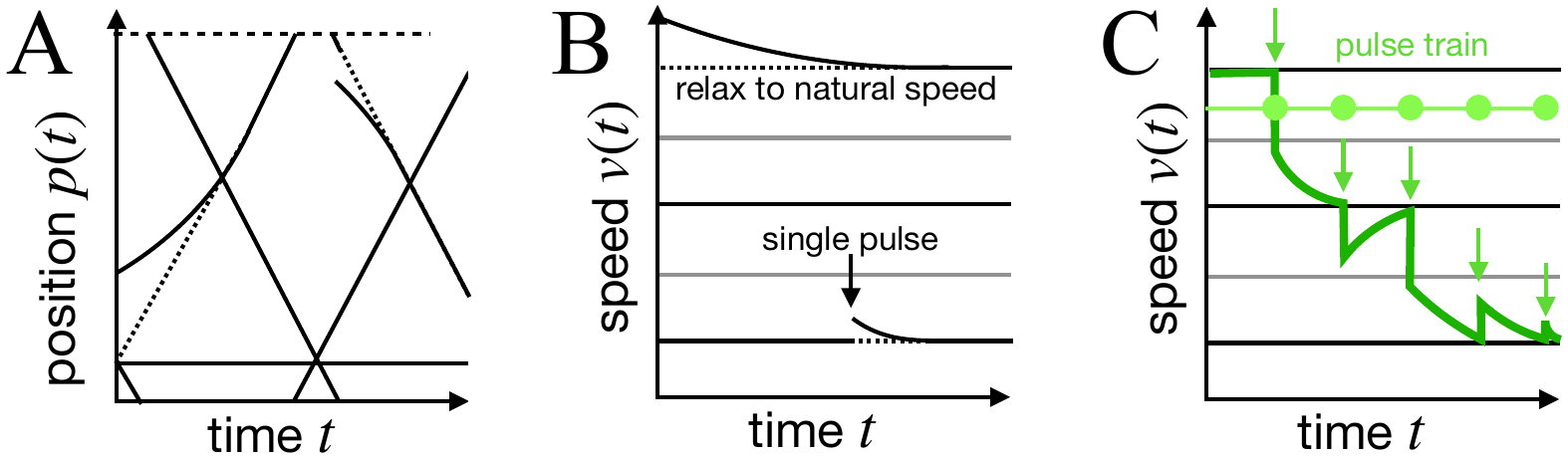} \end{center}
\caption{\textbf{Summary of delay-induced speed selection under stroboscopic inputs.}
\textbf{(A)} Reduced bump dynamics in terms of position $p(t)$. Delayed coupling generates multiple admissible traveling solutions, visible as linear trajectories with distinct slopes corresponding to intrinsic wave speeds. Dotted lines denote natural, stable trajectories; solid lines indicate unstable or transient solutions.
\textbf{(B)} Speed dynamics under isolated perturbations. The mean bump speed $v(t)=\left[p(t)-p(t-\tau)\right]/\tau$ relaxes toward a natural intrinsic speed in the absence of sustained forcing. Black and gray horizontal lines denote stable and unstable speeds, respectively; a single pulse produces a transient deviation followed by relaxation.
\textbf{(C)} Speed dynamics under pulsatile (stroboscopic) forcing. The green horizontal line marks the position of the traveling stimulus, while green dots indicate pulse inputs to the neural field. Repeated pulses induce phase resets (arrows), preventing full relaxation and driving transitions between intrinsic speed branches. The resulting long-term speed can differ from the stimulus speed and may reverse sign, producing stroboscopic motion aliasing.}
\label{fig:scheme}
\end{figure}

Taken together, these observations reveal a hybrid continuous--discrete mechanism for motion
selection.
Delayed neural interactions create a lattice of admissible traveling speeds, while impulsive
inputs act as discrete perturbations that can move the system between neighboring basins of
attraction.
Whether the bump ultimately tracks the stimulus or converges to a different intrinsic speed
depends on how these impulses interact with the underlying delay-induced dynamics.
Figure~\ref{fig:scheme} provides a summary of how impulsive forcing interacts with the
delay-induced speed lattice in the reduced bump dynamics.
In the absence of external input, delayed coupling organizes the autonomous dynamics into a
discrete set of admissible traveling solutions with distinct propagation speeds
(Fig.~\ref{fig:scheme}A), corresponding to stable and unstable branches of $p(t)$.
An isolated impulse perturbs the system within this phase-space structure:
small perturbations decay back to the same intrinsic speed, while sufficiently strong impulses
can push the state across a separatrix and trigger a transition to a neighboring speed branch
(Fig.~\ref{fig:scheme}B).
Under pulsatile (stroboscopic) forcing, repeated phase resets interrupt relaxation and promote
systematic branch switching, so that the dynamics converge to one of the admissible intrinsic
speeds (Fig.~\ref{fig:scheme}C).
As a result, the long-term selected speed may differ from the stimulus speed and can even
reverse direction, yielding quantized motion and stroboscopic aliasing.

While the analysis above assumes a uniform interlayer delay, the underlying mechanism is not tied to this simplification.
Axonal delays can depend on spatial separation, reflecting finite conduction speeds or heterogeneous routing.
In the next section, we extend the interface and phase-reduction framework to
spatially dependent delays $\tau(x-y)$ and show that the combination of delay heterogeneity
and discrete forcing further enriches the speed-selection landscape, while preserving the
delay-induced lattice structure that underlies motion quantization.

\section{Spatially dependent delays}

Thus far, we have assumed a uniform interlayer delay $\tau \equiv \tau_0$, which allowed for a transparent
characterization of delay-induced multistability and speed quantization.
In many neural systems, however, axonal transmission times depend on spatial separation,
reflecting finite conduction velocities, heterogeneous routing, or layer-specific geometry~\cite{atay2004,coombes2007,bojak2010,laing2006}.
Such spatially dependent delays naturally arise in long-range cortical projections and
introduce an additional source of structure into the effective coupling~\cite{swadlow1980,stoelzel2017}.

In this section, we extend the interface and phase-reduction framework to delays of the form
$\tau(x-y)$.
We show that spatially dependent delays deform, but do not destroy, the lattice of admissible traveling speeds identified above.
Instead, delay heterogeneity reshapes the basins of attraction and biases speed selection in a geometry-dependent manner, while preserving the core mechanism by which delayed feedback can stabilize propagating waves in layered neural fields.

We consider a delay-coupled neural field model in which interlayer transmission delays depend explicitly on spatial separation. Allowing axonal propagation times to vary with distance captures finite conduction speeds and heterogeneous inter-areal geometry~\cite{coombes2003waves,atay2004}:
\begin{equation}\label{dubdelay_spatial}
\pd_t u_j= -u_j+ w*f(u_j)
+ \int_{-\pi}^{\pi} w_c(x-y)\, f\!\big(u_{k}(y,t-\tau(x-y))\big)\,dy
+ I_j(x,t),
\end{equation}
where $j,k\in\{1,2\},\ k\neq j$. Here $\tau(z)$ is a $2\pi$-periodic delay kernel defined on $z=x-y$ modulo $2\pi$ and restricted to $[-\pi,\pi]$, so that $|z|$ represents geodesic distance on the ring. The uniform-delay model is recovered as the special case $\tau(z)\equiv\tau_0$. A canonical example is a finite-conduction delay, $\tau(z)=\tau_0+\frac{d(z)}{v}$, where $d(z)$ denotes geodesic distance and $v$ is an effective propagation speed~\cite{coombes2003waves,hutt2003}.

\subsection{Traveling bumps under spatially dependent delays}
For stationary solutions $u_j(x,t)=U_j(x)$, the delay terms are time-independent, so the existence of stationary bumps is unchanged from the uniform-delay case. We therefore seek uniformly translating bump solutions of the form
\[
u_1(x,t)=U_1(\xi),\qquad u_2(x,t)=U_2(\xi-s),\qquad \xi=x-ct,
\]
where $c$ is the wave speed and $s$ is a constant interlayer phase offset ($s=0$ corresponds to co-located bumps). Substituting into \eqref{dubdelay_spatial} and writing $z=x-y$ yields
\begin{subequations}\label{eq:TW_spatialdelay_general}
\begin{align}
-c\,U_1'(\xi) &= -U_1(\xi) + (w*f(U_1))(\xi) + \mathcal{C}_{12}[U_2](\xi),\\
-c\,U_2'(\xi-s) &= -U_2(\xi-s) + (w*f(U_2))(\xi-s) + \mathcal{C}_{21}[U_1](\xi-s).
\end{align}
\end{subequations}
where the delayed interlayer coupling operators are
\begin{align*}
\mathcal{C}_{12}[U](\xi) &\equiv \int_{-\pi}^{\pi} w_c(z)\,
f\!\Big(U(\xi - z + c\,\tau(z)-s)\Big)\,dz, \\
\mathcal{C}_{21}[U](\xi) &\equiv \int_{-\pi}^{\pi} w_c(z)\,
f\!\Big(U(\xi - z + c\,\tau(z))\Big)\,dz.
\end{align*}
Spatially dependent delays therefore appear as a separation-dependent \emph{phase advance}: inputs from offset $z$ act at the comoving location $\xi-z+c\tau(z)$.

We assume single-bump solutions with Heaviside firing rates $f(u)=H(u-\theta)$. Let $p(t)$ denote the center of the layer-1 bump, so that its active region is $(p(t)-a_1,p(t)+a_1)$, and assume the layer-2 bump is shifted by $s$, with active region $(p(t)+s-a_2,p(t)+s+a_2)$. For uniformly translating solutions, $p(t)=ct$, and in the comoving coordinate the active intervals are $(-a_1,a_1)$ and $(s-a_2,s+a_2)$. The threshold conditions are therefore
\[
U_1(\pm a_1)=\theta,\qquad U_2(s\pm a_2)=\theta.
\]

Evaluating \eqref{eq:TW_spatialdelay_general} at the bump edges yields traveling-wave self-consistency conditions. For Heaviside firing rates, delayed synaptic input contributes to the threshold only when the corresponding comoving location lies within the active interval of the opposing layer. This gives
\begin{align*}
\theta &= \int_{-a_1}^{a_1} w(a_1-y)\,dy
+ \int_{-\pi}^{\pi} w_c(z)\,
\mathbf 1_{[\,s-a_2,\,s+a_2\,]}\!\big(a_1-z+c\tau(z)\big)\,dz,\\
\theta &= \int_{-a_2}^{a_2} w(a_2-y)\,dy
+ \int_{-\pi}^{\pi} w_c(z)\,
\mathbf 1_{[\, -a_1,\, a_1\,]}\!\big(s+a_2-z+c\tau(z)\big)\,dz,
\end{align*}
where $\mathbf 1_{[A,B]}$ denotes the indicator function. These conditions make explicit the geometric role of spatially dependent delays: for each offset $z$, delayed input contributes only if the phase-shifted location $\xi-z+c\tau(z)$ lies within the active region of the opposing layer. In the uniform-delay limit $\tau(z)\equiv\tau_0$, this reduces to a rigid shift by $c\tau_0$, while heterogeneous delays deform this inclusion region in a speed- and geometry-dependent manner. To obtain explicit speed equations, we now specialize to cosine connectivity and smooth bump profiles. As in the uniform-delay case, this yields closed-form algebraic conditions, with the uniform-delay factors $\cos(c\tau_0)$ and $\sin(c\tau_0)$ replaced by geometry-dependent averages of $\cos(c\tau(z))$ and $\sin(c\tau(z))$.

\subsection{Explicit reduction for cosine weights}
To make the effect of spatially dependent delays explicit, we now specialize to cosine intra- and inter-layer connectivity \eqref{w}. As in the uniform-delay case, this choice closes the convolution operators on the first Fourier mode and allows traveling bump solutions to be represented exactly by sinusoidal profiles,
\[
U_1(\xi)=A\cos\xi, \qquad U_2(\xi-s)=B\cos(\xi-s),
\]
with half-widths $a_1,a_2$ determined by the threshold conditions
$A\cos a_1=\theta$ and $B\cos a_2=\theta$.

Remarkably, introducing spatially dependent delays does not alter the algebraic structure of the traveling-wave reduction. Instead, all effects of delay heterogeneity are captured by a single complex-valued coefficient,
\begin{equation}\label{eq:Kdef}
K(c)=\frac{1}{\pi}\int_{-\pi}^{\pi}\cos z\,e^{ic\tau(z)}\,dz
= C(c)+iS(c),
\end{equation}
which encodes both a geometry-dependent gain and an effective phase shift through its real and imaginary parts,
\[
C(c)=\frac{1}{\pi}\int_{-\pi}^{\pi}\cos z\,\cos\!\big(c\tau(z)\big)\,dz,
\qquad
S(c)=\frac{1}{\pi}\int_{-\pi}^{\pi}\cos z\,\sin\!\big(c\tau(z)\big)\,dz.
\]
In the uniform-delay limit $\tau(z)\equiv\tau_0$, this coefficient reduces to $K(c)=e^{ic\tau_0}$, recovering the speed-selection equations derived previously.

We focus first on the co-located branch $s=0$ with equal half-widths $a_1=a_2=a$. In this case, the convolution identities for cosine weights yield the same amplitude and speed relations as in the uniform-delay setting, with the uniform phase factors $\cos(c\tau_0)$ and $\sin(c\tau_0)$ replaced by the geometry-dependent averages $C(c)$ and $S(c)$:
\begin{align*}
A = 2\sin(a)\big(1+\bar w\,C(c)\big),
\qquad
cA = -2\bar w \sin(a)\,S(c).
\end{align*}
Eliminating $A$ produces the implicit speed selection equation
\begin{equation}\label{eq:speedeq_spatialdelay}
f_\tau(c)\equiv c + c\,\bar w\,C(c) + \bar w\,S(c)=0,
\end{equation}
while the corresponding half-width follows from the threshold condition as
\begin{equation}\label{eq:width_spatialdelay}
\theta=\sin(2a)\big(1+\bar w\,C(c)\big).
\end{equation}
Together, \eqref{eq:speedeq_spatialdelay} and \eqref{eq:width_spatialdelay} define a
discrete set of admissible traveling speeds, forming a geometry-dependent deformation of the uniform-delay speed lattice governed entirely by the oscillatory averages $C(c)$ and
$S(c)$.

We now illustrate this reduction for two representative classes of spatial delay profiles, both of which admit closed-form expressions for $C(c)$ and $S(c)$.

\paragraph{Finite-conduction delays}
For $\tau(z)=\tau_0+|z|/v$, where $|z|$ denotes the geodesic distance on the ring, the delay profile is even and the kernel \eqref{eq:Kdef} simplifies to
\[
K(c)=\frac{2}{\pi}e^{ic\tau_0}\int_{0}^{\pi}\cos z\,e^{i\alpha z}\,dz,
\qquad
\alpha=\frac{c}{v}.
\]
Evaluating this integral yields
\[
K(c)=\frac{2\,e^{ic\tau_0}}{\pi}
\left[
\frac{\sin(\alpha\pi)}{\alpha^2-1}
- i\,\frac{\alpha\big(1+\cos(\alpha\pi)\big)}{\alpha^2-1}
\right],
\]
providing explicit expressions for $C(c)$ and $S(c)$ and allowing admissible traveling speeds to be computed directly from \eqref{eq:speedeq_spatialdelay}.

\paragraph{Smooth nonnegative delay heterogeneity}
As an alternative that preserves $\tau(z)\ge\tau_0$, we consider
\[
\tau(z)=\tau_0+\frac{1-\cos z}{v}.
\]
Writing $\beta=c/v$, the delay factor can be expressed as
\[
e^{ic\tau(z)}=e^{ic(\tau_0+1/v)}e^{-i\beta\cos z},
\]
and substitution into \eqref{eq:Kdef} gives
\begin{equation}\label{eq:K_1minuscos_closed}
K(c)=-2i\,e^{ic(\tau_0+1/v)}\,J_1\!\left(\frac{c}{v}\right).
\end{equation}
Defining $\psi=c(\tau_0+1/v)$, this yields
\[
C(c)=2J_1\sin\psi,
\qquad
S(c)=-2J_1\cos\psi,
\]
and substituting into \eqref{eq:speedeq_spatialdelay} gives the explicit speed selection condition
\[
c + 2\bar w\,J_1\!\left(\frac{c}{v}\right)\big(c\sin\psi-\cos\psi\big)=0,
\]
with the corresponding half-width
\[
\theta=\sin(2a)\big(1+2\bar w\,J_1\!\left(\frac{c}{v}\right)\sin\psi\big).
\]
In both examples, spatially dependent delays leave the underlying algebraic structure of traveling bump solutions intact while reshaping the admissible speed set through geometry-dependent gain and phase modulation.

\subsection{Stroboscopic forcing with spatially dependent delays}

We extend the stroboscopic forcing analysis to spatially dependent interlayer delays $\tau(z)$. As in the uniform-delay setting, we assume a separable input
\[
I(x,t)=I_t(t)\,I_\xi(x),
\]
and allow the bump position to evolve under the combined influence of delayed coupling and intermittent forcing. We assume the interlayer coupling kernel $w_c(z)$ and delay profile $\tau(z)$ are even, and restrict attention to phase-aligned solutions in which both layers share a common bump profile and a single time-dependent position,
\[
u_j(x,t)=U(x-p(t)),\qquad j=1,2,
\]
with $p(t)$ denoting the bump center.

Substituting into \eqref{dubdelay_spatial} and working in the comoving coordinate $\xi=x-p(t)$ yields
\begin{align}
-p'(t)U'(\xi)=\;&-U(\xi)+(w*f(U))(\xi) \nonumber\\
&+\int_{-\pi}^{\pi} w_c(z)\, f\!\big(U(\xi-z+p(t)-p(t-\tau(z)))\big)\,dz +I_t(t)I_\xi(\xi).
\end{align}
Thus, spatially dependent delays enter the reduced dynamics through the phase difference $p(t)-p(t-\tau(z))$.

Specializing to cosine interlayer connectivity $w_c(z)=\cos z$ and cosine input $I_\xi(\xi)=\cos\xi$, and projecting onto the first Fourier mode as in Section~\ref{sect:strobconst}, the bump dynamics reduce to a scalar delay differential equation for $p(t)$. Evaluating at the interfaces $\xi=\pm a$ and enforcing
$U(\pm a)=\theta$ yields
\begin{equation}
p'(t)=
-\frac{2\bar w \sin(a)}{\theta}
\int_{-\pi}^{\pi}\cos z\,
\sin\!\big(p(t)-p(t-\tau(z))\big)\,dz
-\frac{I_t(t)}{\theta}\sin\!\big(p(t)-ct\big)\cos(a).
\label{eq:p_delay_spatial}
\end{equation}
For even $\tau(z)$, the delayed coupling term may equivalently be written as $2\int_{0}^{\pi}\cdots\,dz$.

In the absence of forcing ($I_t\equiv0$), uniformly translating solutions $p(t)=ct$ reduce the coupling term to a $z$-integral involving $\sin(c\,\tau(z))$, recovering the intrinsic speed selection mechanism analyzed above.

To isolate the effects of discrete temporal sampling, we specialize to impulsive forcing
\[
I_t(t)=\sum_{n=0}^\infty A\,\delta(t-t_n).
\]
Integrating \eqref{eq:p_delay_spatial} across an impulse at $t=t_n$ yields the jump map
\begin{equation}
p(t_n^+)-p(t_n^-)
=-\frac{A}{\theta}\sin\!\big(p(t_n^-)-ct_n\big)\cos(a),
\label{eq:jump_spatial}
\end{equation}
which is identical to the uniform-delay case; spatially dependent delays influence the dynamics solely through the inter-pulse evolution.

\begin{figure}[t!]
\begin{center}
\includegraphics[width=0.85\textwidth]{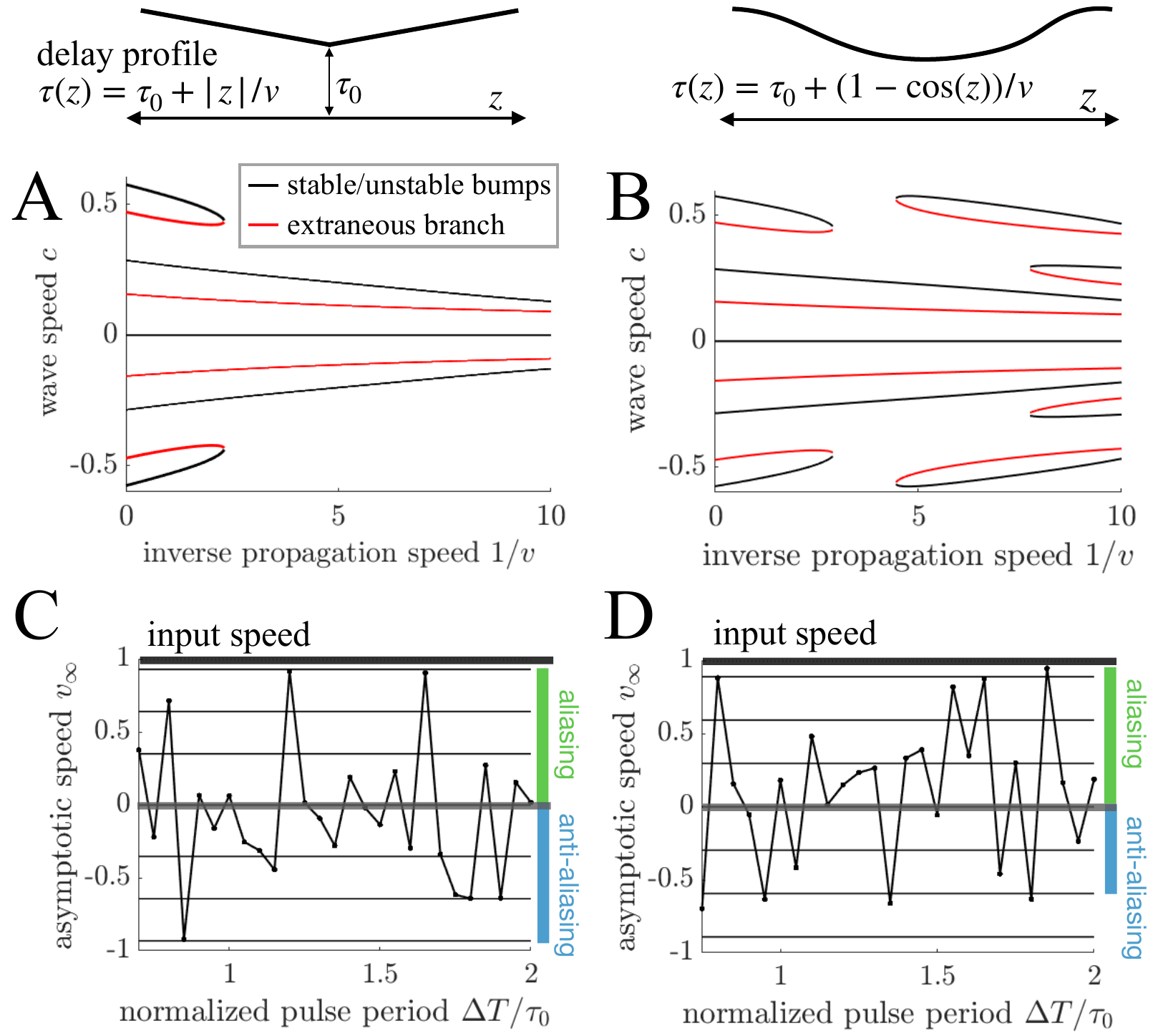}
\end{center}
\caption{Effect of spatially dependent delays on intrinsic traveling speeds and stroboscopic entrainment.
\textbf{(A)} Admissible traveling wave speeds $c$ as a function of inverse propagation speed $1/v$ for finite-conduction delays $\tau(z)=\tau_0+|z|/v$.
\textbf{(B)} Corresponding speed structure for periodic delays $\tau(z)=\tau_0+(1-\cos z)/v$. Black curves denote admissible traveling bump solutions satisfying the width condition, while red curves indicate extraneous branches that do not correspond to physically realizable bumps. Parameters are $\tau_0 = 20$ and $\bar{w} = 1$.
\textbf{(C--D)} Long-time average bump speed $v_\infty$ under pulsatile forcing as a function of the normalized pulse period $\Delta T/\tau_0$, for the delay profiles in \textbf{(A)} and \textbf{(B)}, respectively. Horizontal black lines indicate admissible intrinsic speeds, organizing regions of aliasing ($0<v<c_{\rm in}$) and anti-aliasing ($v<0$) in the stimulus-driven response. Additional parameters are $v = 1$, $A = 3$, and $c_{\rm in} = 1$.}
\label{fig:spatial_delay_speeds}
\end{figure}

The analytical speed selection equation~\eqref{eq:speedeq_spatialdelay} predicts that spatially dependent delays deform, but do not eliminate, the discrete lattice of admissible traveling speeds generated by delayed interlayer coupling. Figure~\ref{fig:spatial_delay_speeds} illustrates this deformation for two representative classes of delay profiles and shows how the resulting intrinsic speed structure organizes the response to pulsatile forcing. Figures~\ref{fig:spatial_delay_speeds}A--B demonstrate how spatially structured delays reshape the intrinsic traveling speeds of the delay-coupled neural field. For finite-conduction delays of the form $\tau(z)=\tau_0+|z|/v$, admissible speed branches bend and compress as the inverse propagation speed $1/v$ increases, reflecting the growing influence of distance-dependent phase shifts. For periodic delays $\tau(z)=\tau_0+(1-\cos z)/v$, the geometry of the delay kernel can induce branch splitting and the appearance of additional admissible speeds. In both cases, admissible traveling bumps persist as isolated branches in speed--parameter space, while extraneous branches correspond to solutions that fail to satisfy the bump width condition and are therefore not physically realizable. The persistence of a discrete speed set across delay profiles indicates that speed quantization is a robust consequence of delayed coupling rather than an artifact of uniform delays.

Figures~\ref{fig:spatial_delay_speeds}C--D show the long-time average bump speed $v_\infty$ under pulsatile forcing for the two spatial delay profiles. Horizontal black lines indicate admissible intrinsic speeds predicted by the autonomous spatial-delay dynamics, which organize the stimulus-driven response into distinct locking regions. Depending on the pulse period relative to the baseline delay, the bump converges to an intrinsic speed that may be slower than the stimulus (aliasing) or opposite in sign (motion reversal). Between impulses, the bump evolves according to the autonomous dynamics obtained from~\eqref{eq:p_delay_spatial} with $I_t\equiv0$, while impulses induce instantaneous phase shifts that move the system between basins of attraction associated with different intrinsic speeds. Spatial delay heterogeneity biases the location and width of these locking regions but does not alter the underlying mechanism: delayed feedback generates a geometry-dependent lattice of admissible speeds, and pulsatile inputs select among them through discrete phase resets, yielding quantized long-term motion even under realistic, distance-dependent transmission delays.

\section{Discussion}
\label{sec:conclusions}

We developed a delay-coupled neural field model of visual motion representation and showed that transmission delays fundamentally restructure the dynamics of bump attractors. Both uniform and spatially structured delays generate a discrete lattice of admissible traveling bump speeds, yielding multiple stable motion representations even in the absence of input. Spatially dependent delays deform this speed lattice in a geometry-dependent manner, biasing speed selection while preserving the underlying multistability. Under continuously rotating stimuli, sufficiently strong inputs suppress this intrinsic competition and stabilize a uniquely traveling bump that entrains to the stimulus with a fixed phase lag. In contrast, under stroboscopic forcing, delayed coupling and impulsive inputs interact to produce quantized speed selection and robust transitions between distinct traveling states, including reversals of motion direction. Together, these results identify delayed interlayer feedback as a minimal circuit mechanism capable of generating motion aliasing and wagon-wheel-like percepts without invoking an explicit external sampling clock.

A central and novel finding of this work is that delayed interlayer coupling alone is sufficient to generate a discrete family of coexisting traveling wave solutions with distinct, stable propagation speeds. Unlike classical neural field models, where traveling waves typically form continuous families parameterized by asymmetry or input strength, the present system admits intrinsically quantized speeds. This multistability emerges in a simple, spatially homogeneous architecture with finite transmission delays, without requiring network heterogeneity, external forcing, or explicit discretization. The resulting speed lattice admits a low-dimensional analytical characterization, revealing saddle-node structure and alternating stability across branches. To our knowledge, this organization of multistable traveling wave speeds has no direct analogue in prior neural field models and represents a qualitatively new dynamical consequence of delayed coupling in spatially extended systems.

Delays have long been known to influence neural field dynamics by shaping wave propagation, pattern formation, and stability through finite transmission speeds and delayed feedback \cite{atay2004,hutt2003,laing2006,coombes2007}. Prior work has shown that delays can modify wave speeds, induce oscillatory instabilities, or support coherent traveling activity. The present results extend this literature by showing that delayed coupling can qualitatively restructure the solution set itself, organizing traveling bumps into a discrete family of coexisting stable propagation speeds rather than a continuous branch. This speed quantization arises from the interaction of delay and recurrent coupling and persists under spatially structured delays, a phenomenon that has not been explicitly characterized in existing delayed neural field models.

Several extensions of this work are natural. Incorporating stochasticity would allow a systematic investigation of noise-induced bump motion and transitions through asymptotic reductions to stochastic delay equations~\cite{kilpatrick2013wandering,kilpatrick2015delay}. Such an approach motivates further study of the associated linear operators, including solvability conditions arising from the Fredholm alternative, which in the presence of delays likely require the framework of functional differential equations and appropriate bilinear forms~\cite{hale2013,kotani2012}. In addition, while we focused here on constant-amplitude inputs and impulsive forcing, alternative protocols such as sinusoidally modulated or square-wave inputs may align more closely with leading theories of perceptual cycles and temporal sampling in vision~\cite{vanrullen2016perceptual}. Analyzing such inputs will require a more refined treatment of entrainment in delayed systems under nonlinear forcing.

Finite conduction delays in recurrent cortical networks are known to generate traveling waves that shape excitability and support short-term prediction in visual cortex \cite{davis2021,benigno2023}. More broadly, growing experimental evidence suggests that traveling waves play a role in information transfer, prediction, and coordination across sensory and associative cortices \cite{friston2019,luo2025}. The present work complements this literature by focusing on a reduced neural field framework in which the consequences of delayed coupling can be characterized analytically at the level of coherent structures. Whereas prior studies emphasize wave emergence and predictive modulation, our results show that delayed interlayer feedback can organize traveling solutions into a discrete set of coexisting stable propagation speeds. These speeds can be selectively stabilized or switched through continuous or stroboscopic forcing, providing a distinct dynamical contribution to the theory of cortical wave entrainment and control.

In summary, we employed and extended analytical techniques for neural field models to study a delay-coupled system representing populations encoding angular motion in the visual field. Within this framework, bump solutions emerge as coherent representations of perceived stimulus location and can be stabilized, entrained, or destabilized by time-dependent inputs. The analysis further reveals how delayed feedback can induce robust propagation opposite to the driving stimulus, providing a mechanistic realization of the wagon wheel illusion within a continuous dynamical system. More broadly, these results demonstrate how delayed recurrent interactions can endow neural circuits with rich and counterintuitive motion representations through intrinsically generated multistability and hybrid entrainment dynamics.

\section{Numerical Methods}
\label{sec:numerics}

Simulations were performed on the periodic domain $x\in[-\pi,\pi)$ using a uniform grid with $\Delta x=2\pi/1000$ and forward Euler time stepping with $\Delta t=10^{-3}$. Spatial convolution integrals were evaluated using the composite trapezoidal rule. Delayed terms were computed from stored solution history via linear interpolation in time, ensuring consistent treatment of spatially dependent delays. Impulsive forcing was implemented as instantaneous jump updates at pulse times, consistent with the derived jump conditions. Nonlinear self-consistency, speed-selection, and Evans function equations were solved in MATLAB using \texttt{fzero} and \texttt{fsolve}.

\end{document}